\documentclass[iop,apj]{emulateapj}

\usepackage{apjfonts}
\usepackage{multirow}
\usepackage{rotating}
\usepackage{amssymb,amsmath}
\usepackage{hyperref}
\usepackage{longtable}
\usepackage{color,soul}

\def\hst{{\it HST}}

\newcommand{\lta}{\lesssim}
\newcommand{\gta}{\gtrsim}
\newcommand{\kms}{\>{\rm km}\,{\rm s}^{-1}}
\newcommand{\masyr}{\>{\rm mas}\,{\rm yr}^{-1}}
\newcommand{\kpc}{\>{\rm kpc}}

\newcommand{\muw}{\mu_{W}}
\newcommand{\mun}{\mu_{N}}

\shorttitle{HST Proper Motions along the Sagittarius Stellar Stream: I}
\shortauthors{Sohn et al.}

\begin{document}

\title{Hubble Space Telescope Proper Motions Along the Sagittarius
  Stream: \\I. Observations and Results for Stars in Four Fields}

\author{
Sangmo Tony Sohn\altaffilmark{1,2}, 
Roeland P. van der Marel\altaffilmark{1}, 
Jeffrey L. Carlin\altaffilmark{3},
Steven R. Majewski\altaffilmark{4},
\\
Nitya Kallivayalil\altaffilmark{4},
David R. Law\altaffilmark{1},
Jay Anderson\altaffilmark{1},
and  
Michael H. Siegel\altaffilmark{5}}

\altaffiltext{1}{Space Telescope Science Institute, 
                 3700 San Martin Drive, Baltimore, MD 21218, USA}
\altaffiltext{2}{Department of Physics and Astronomy, 
                 The Johns Hopkins University, Baltimore, MD 21218, USA}
\altaffiltext{3}{Department of Physics, Applied Physics and Astronomy,
                 Rensselaer Polytechnic Institute,
                 110 8th Street, Troy, NY 12180, USA}
\altaffiltext{4}{Department of Astronomy, 
                 University of Virginia 
                 Charlottesville, VA 22904-4325, USA}
\altaffiltext{5}{Department of Astronomy, 
                 Pennsylvania State University,
                 525 Davey Laboratory, University Park, PA 16802, USA}

\email{tsohn@jhu.edu}

\begin{abstract}
We present a multi-epoch Hubble Space Telescope (\hst) study of
stellar proper motions (PMs) for four fields spanning 200 degrees
along the Sagittarius (Sgr) stream: one trailing arm field, one field
near the Sgr dwarf spheroidal tidal radius, and two leading arm
fields. We determine absolute PMs of dozens of individual stars per
field, using established techniques that use distant background
galaxies as stationary reference frame. Stream stars are identified
based on combined color-magnitude diagram and PM information. The
results are broadly consistent with the few existing PM measurements
for the Sgr galaxy and the trailing arm.  However, our new results
provide the highest PM accuracy for the stream to date, the first PM
measurements for the leading arm, and the first PM measurements for
individual stream stars; we also serendipitously determine the PM of
the globular cluster NGC~6652. In the trailing-arm field, the
individual PMs allow us to kinematically separate trailing-arm stars
from leading-arm stars that are 360 degrees further ahead in their
orbit.  Also, in three of our fields we find indications that two
distinct kinematical components may exist within the same arm and wrap
of the stream. Qualitative comparison of the
\hst\ data to the predictions of the Law \& Majewski and Pe\~narrubia
et al. $N$-body models show that the PM measurements closely follow
the predicted trend with Sgr longitude. This provides a successful
consistency check on the PM measurements, as well as on these N-body
approaches (which were not tailored to fit any PM data).
\end{abstract}

\keywords{Astrometry ---
galaxies: kinematics and dynamics --- 
Local Group}

\section{Introduction}
\label{sec:intro}

With the advent of progressively deeper photometric surveys, it has
become apparent that galactic halos are threaded with the phase-mixed
debris of multiple generations of dwarf satellites that have been
destroyed by the tides of their host's gravitational potential.
Surveys have found many streams in the MW, possibly constituting the
primary source of the Galactic stellar halo \citep[see, e.g., the
  review of][]{hel08}. Streams have been found in other nearby
galaxies as well, including, e.g., M31 \citep{fer02} and NGC 4449
\citep{mar12}. The first known nearby stream, the Magellanic stream,
has been detected only in H\textsc{i}. Its exact origins continue to
be the topic of intense debate \citep[e.g.,][]{nid08,bes10}. 
By comparison, the origin of the brightest and most prominent stellar 
stream around the MW, the Sgr stream, is better understood. It emanates 
from the Sgr dSph first detected by \citet{iba94}, and its lengthy 
tidal features wrap entirely around the MW. This makes the Sgr stream 
an ideal target for probing in detail several important topics, 
including the tidal disruption of dwarf galaxies, the hierarchical 
buildup of stellar halos, and the shape, orientation and mass of the 
MW's dark halo.

Much observational effort has been put into the detection and
characterization of the stellar stream from the Sgr dSph. Early
observations included measurements of high-latitude carbon stars 
\citep{iba01} and various pencil-beam surveys \citep{mat98,maj99,
mar01,mar02}. In the past decade, however, our understanding of the 
scope and significance of the stream has been revolutionized by the 
deep, wide-field views provided by the Two Micron All-Sky Survey 
(2MASS) and Sloan Digital Sky Survey (SDSS). The 2MASS survey 
revealed a large population of young, relatively metal-rich M-giant 
stars that belong to the Sgr stream, spanning across the entire sky
\citep{maj03}. SDSS observations showed that the debris stream 
leading Sgr along its orbit continues to be well-defined through 
the North Galactic Cap, as it passes over the
solar neighborhood towards the Galactic anticenter \citep{bel06}. 
In addition to its location and width on the sky, there is now also 
a wealth of other data available for the stream. For example, CMDs 
have provided the variation of distance along the stream, and 
spectroscopy has provided the variation of the mean line-of-sight 
(LOS) velocity along the stream, as well as the velocity dispersion 
of the debris \citep{maj04,bel06,mon07,car12,kop12,sla13,bel14}.

The Sgr stream has been the topic of intense modeling efforts over the
past decade. While the models have been able to successfully reproduce
many features of the stream, their ability to constrain the shape of
the MW's dark halo continues to be hotly debated. Depending on which
data sets were fitted, claims have been made in favor of an oblate halo
\citep{joh05,mar07}, an approximately spherical halo
\citep{iba01,fel06}, and a prolate halo \citep{hel04}. Most recently,
\citet{law09}, \citet[][hereafter, LM10]{law10}, and \citet{deg13}
demonstrated that one obtains the best fit to almost all available
data if one allows for a triaxial halo. However, this then implies a
rather unexpected halo orientation. The best-fit halo in the LM10
model is near-oblate, but with the symmetry axis perpendicular to the
symmetry axis of the MW disk, and pointing at the Galactic Center.
Of course, as recognized by LM10 this does not necessarily imply that 
the halo must be triaxial this way. If some of the assumptions of the 
model were to be relaxed, it is possible that other models might 
provide better fits \citep[see also][]{ver13,iba13}. The implied MW 
halo shape is not the only open question about the Sgr stream. For 
example, the observed bifurcation of the stream on the leading side 
\citep{bel06} is not understood, yet.

To further clarify this situation, it is important to also have access
to proper motions (PMs) of stream stars. Some limited PM data now 
indeed exist. Several studies have measured the average PM of the Sgr 
dSph itself \citep{din05,pry10,mas13}.  Recently, the first
measurements of the average PM in fields in the trailing arm of the
stream have been reported \citep{car12,kop13}. This is a very exciting
development, since addition of the two transverse components of motion
yields fully six-dimensional phase-space information. Such information
makes many model degeneracies that otherwise exist disappear. With
PMs, it also becomes possible to constrain the velocity of the Sun in
the MW disk \citep{car12}. However, there is significant room for
improvement in the available PM observations. First, smaller error 
bars than have been reported thus far would improve the discriminatory
power of the PM data. Second, measurements in the leading arm would
help constrain MW halo properties over a wider range of radii. And
third, measurements for individual stars, as opposed to average PMs,
would allow separation of different stream components and wraps.

In recent years, we have pioneered techniques to obtain extremely
accurate absolute PM measurements from multi-epoch \hst\ imaging,
using distant background-galaxies to define a stationary reference
frame. We have used these techniques to obtain the first ever bulk PM
measurements for the Local Group galaxies M31 \citep{soh12} and Leo~I
\citep{soh13}. Moreover, we have shown that these same techniques can
be used to determine the PMs of individual stars in the MW halo
\citep{dea13}. Here we apply the techniques to new
\hst\ imaging for four fields in the Sgr Stream. These studies are all
part of, and use techniques developed in the context of, the HSTPROMO
collaboration: a set of \hst\ projects aimed at improving our dynamical 
understanding of stars, clusters, and galaxies in the nearby Universe 
through measurement and interpretation of PMs \citep[e.g.,][]{vdm13}.
\footnote{For details see the HSTPROMO home page at
\url{http://www.stsci.edu/~marel/hstpromo.html}}

This paper is organized as follows. In Section~\ref{sec:data}, we
describe the data used for this study, as well as the determination of
photometric and PM measurements for individual stars in the four
target fields: one field in the trailing arm, one field near the Sgr
dSph tidal radius\footnote{The tidal radius is defined here as the 
radius inside which material is bound to the Sgr dSph (see 
Section~\ref{sec:dSphPM}).}, and two fields in the leading arm. These 
fields span a range of more than 200 degrees along the Sgr stream. 
In Section~\ref{sec:sgrstream}, we describe the identification of Sgr 
stream stars in each of the fields, based on the combined CMD and PM 
information. In Section~\ref{sec:pmvar} we discuss the inferred PM 
variation along the stream, and we compare the results to both 
existing PM measurements and to the PM predictions of the LM10 model. 
In Section~\ref{sec:conclusions}, we summarize the main results of 
the paper. An Appendix discusses some details about our modeling of 
point-spread function (PSF) variations between epochs, as well as 
the first measurement of the absolute PM of globular cluster 
NGC~6652.

This is the first paper in a series of two. In Paper II (van der Marel
et al., in prep.) we quantitatively compare the new \hst\ data to Sgr
stream models. We use this comparison to shed new light on topics such
as the structure and distance of the Sgr stream, the solar velocity in
the MW disk, and the shape of the MW's dark halo.

\section{Observations and Data Analysis}
\label{sec:data}

\subsection{First-Epoch Data}
\label{sec:1stepdata}

\begin{figure*}
\epsscale{1.19}
\plotone{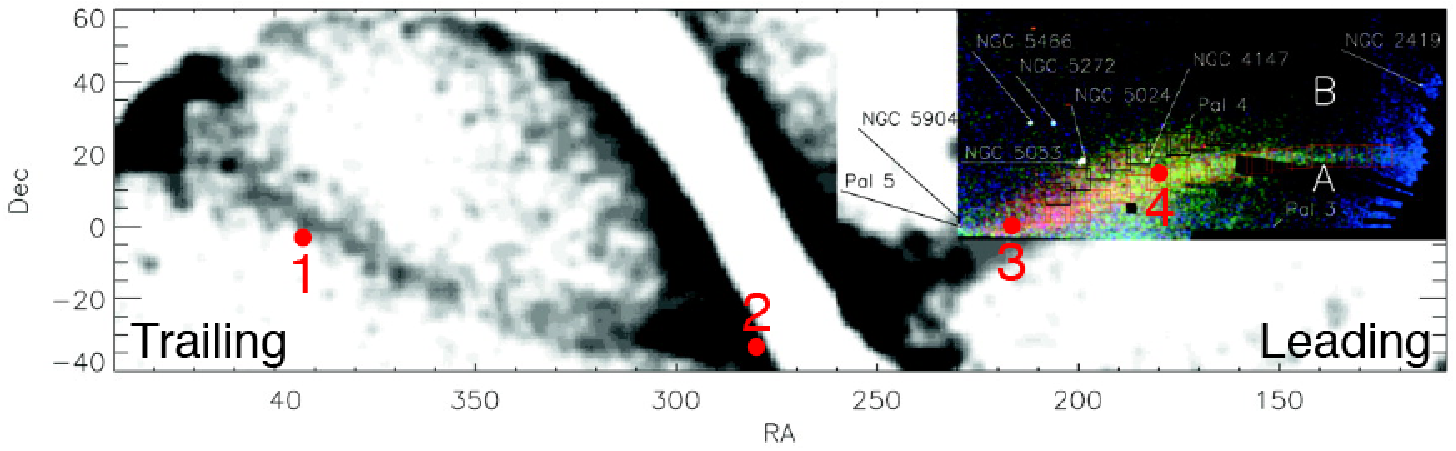}
\epsscale{1.172}
\plotone{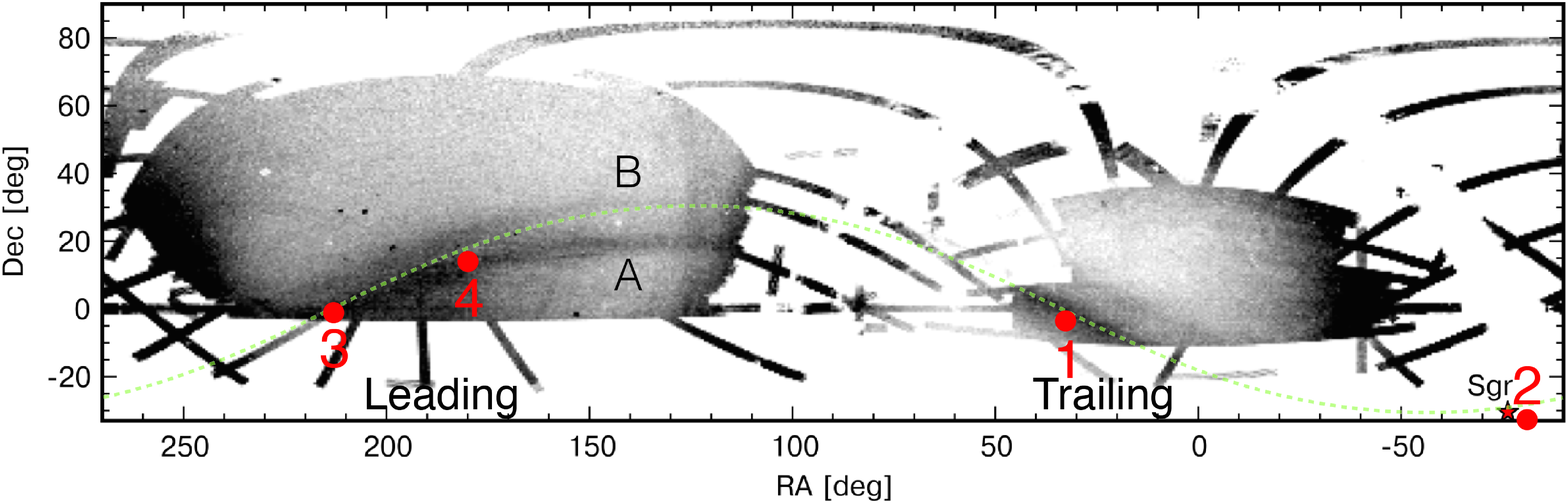}
\caption{Two views of the Sagittarius stream in equatorial coordinates, 
  as it circles the celestial sphere. Top panel 
  \citep[Figure 2 of][]{bel06}: The greyscale reflects the density of M 
  giants selected from 2MASS \citep{maj03}, while the color-map reflects 
  the density of SDSS-selected Stream stars. Bottom panel \citep[Figure 
  1a of][]{kop12}: The greyscale reflects the density of
  main-sequence turnoff stars from the SDSS DR8. The green dashed line
  is the projection of the Sagittarius orbital plane as defined by
  \citet{maj03}. The right ascension ranges of the two panels
  are shifted relative to each other by $\sim 180^{\circ}$ (in the top
  panel, the present location of the Sagittarius dwarf spheroidal
  galaxy is in the middle of the RA range; in the bottom panel it is
  near the right edge, marked by a red star). The trailing and leading
  arms of the stream are labeled in each panel. The leading arm
  bifurcates into branches labeled A and B \citep[following the 
  nomenclature of][]{bel06}. In both panels, the locations of our \hst\ 
  target fields are indicated with numbered red dots.
         \label{fig:fieldlocations}
        }
\end{figure*}

Our goal for this project was to measure accurate \hst\ PMs
of stars located at various positions along the stream. To this end,
we searched the \hst\ archive for existing deep imaging
serendipitously located around the densest parts of the Sgr Stream.
We identified and selected 4 fields along the stream. The field
locations are shown in Figure~\ref{fig:fieldlocations}, and their
characteristics are as follows. 

{\bf FIELD 1} is a field located on the trailing arm, which was
observed to study the morphology of galaxies at $z > 1$ in the Gemini
Deep Deep Survey \citep{abr04} in the context of \hst\ program GO-9760 
(PI: R.~Abraham).

{\bf FIELD 2} is centered on the globular cluster NGC~6652, which was
observed in the context of \hst\ program GO-10775 (PI: A.~Sarajedini)
as part of the ACS Survey of Galactic Globular Clusters \citep{sar07}.
This field is located near the tidal radius of the Sgr dSph.  Some 60
Sgr stream star candidates were serendipitously identified as a
background feature in the NGC~6652 CMD by \citet{sie11}. These 
authors identified a total of six bulge clusters that exhibit Sgr 
background features in their CMDs. From this sample we selected the 
NGC~6652 field, because it has the highest number of bright, 
unblended background galaxies in the field that can be used to define 
an absolute astrometric reference frame.

{\bf FIELD 3} and {\bf FIELD 4} are located on the leading arm. The
the two fields are separated by $36\degr$ on the sky. \citet{bel06} 
identified two branches in the leading arm, which are denoted A~and~B 
in Figure~\ref{fig:fieldlocations}. {\bf FIELD 4} lies along the main 
A-branch, while {\bf FIELD 3} lies near the right ascension where the 
two branches bifurcate. The fields were observed by \hst\ to study 
$z \simeq 4$ QSOs identified from the SDSS in the context of program 
GO-10417 (PI: X. Fan). Each field therefore has a bright QSO, which 
provides a point-source astrometric reference that is independent of 
the background galaxies in the field.

All fields were observed with the Wide Field Channel on the Advanced
Camera for Surveys (ACS/WFC). With the exception of {\bf FIELD 2}, all
fields were observed only in a single filter. For {\bf FIELDS 1} and
{\bf 2} the F814W filter was used, and for {\bf FIELDS 3} and {\bf 4}
the F775W filter. Hereafter, we denote both of these filters as
$I$-band, unless there is a need to make a distinction between them.
{\bf FIELD 2} was also observed in F606W.

Table~\ref{tab:obslog} lists the equatorial and Sgr coordinates of the
fields, as well as relevant characteristics of the first-epoch
observations. The Sgr longitude--latitude system $(\Lambda_{\odot},B)$
measures the position relative to a great circle that roughly lies
along the Sgr stream, with the Sgr dSph at $\Lambda_{\odot} = 
0^{\circ}$, as defined by \citet{maj03}.

%
%

\subsection{Second-epoch Data}
\label{sec:2ndepochdata}

The second-epoch data were obtained by us between 2012 April and
October in the context of our science program GO-12564 (PI: R. P. van
der Marel). To enable an optimal astrometric analysis, we used ACS/WFC
with the same filters used in the first-epoch observations. At the
end of each observing sequence for {\bf FIELDS 1}, {\bf 3}, and 
{\bf 4}, we also included short (total exposure times of 
$\sim 2,500$ s per field) $V$-band observations in the F606W filter. 
These enable the construction of CMDs, which are necessary for the 
identification of Sgr stream stars. We matched the orientation and 
field centers of the second-epoch observations as closely as possible 
to those of the first-epoch observations. However, due to 
unavailability of the same guide stars used for the first-epoch 
observations, our second-epoch observations for some of the fields 
required slight differences in the field centers and/or field 
orientations.

Table~\ref{tab:obslog} lists also the relevant characteristics of the
second-epoch observations. The exposure times were chosen to provide 
a sufficient signal-to-noise ratio for a sufficient number of stream
stars and background galaxies in each field to enable accurate PM
determinations. The time baselines between the epochs range
from 6--9 years.

%
\begin{deluxetable*}{lcccccccccc}
\tablecolumns{11}
\tablewidth{0pc}
\tablecaption{\hst\ Target Fields and Observations
              \label{tab:obslog}}
\tablehead{
\colhead{} & \colhead{R.A.} & \colhead{Decl.}  & \colhead{$\Lambda_{\odot}$\tablenotemark{a}} & \colhead{$B_{\odot}$\tablenotemark{a}} & \multicolumn{3}{c}{Epoch 1} & \colhead{} & \multicolumn{2}{c}{Epoch 2 (Prog. ID 12564)} \\
\cline{6-8} \cline{10-11} \\
\colhead{Target} & \colhead{(J2000)} & \colhead{(J2000)} & \colhead{(deg)} & \colhead{(deg)} & \colhead{Prog. ID} & \colhead{Epoch} & \colhead{Exp. Time (s)\tablenotemark{b}} & \colhead{} & \colhead{Epoch} & \colhead{Exp. Time (s)\tablenotemark{b}} 
}
\startdata
{\bf FIELD 1} & 02:09:37.4 & $-$04:38:31.6 & 103.64 & 3.37 & \phn9760 & 2003.83 &    10,503 & & 2012.83 & 7,204 \\
{\bf FIELD 2} & 18:35:45.7 & $-$32:59:24.9 & 356.47 & 4.75 &    10775 & 2006.40 & \phn1,700 & & 2012.29 & 7,390 \\
{\bf FIELD 3} & 14:12:05.7 & $-$01:01:52.5 & 287.05 & 1.21 &    10417 & 2005.37 & \phn7,437 & & 2012.37 & 7,288 \\
{\bf FIELD 4} & 11:59:06.4 & $+$13:37:37.9 & 251.07 & 4.11 &    10417 & 2005.39 & \phn5,024 & & 2012.37 & 6,431
\enddata
\tablenotetext{a}{Coordinates in the Sgr system as defined by \citet{maj03}.\\}
\tablenotetext{b}{Total exposure time of the $I$-band observations used for astrometric analysis.}
\end{deluxetable*}
%

\subsection{Astrometric Measurements}
\label{sec:starastrom}

We compared the two epochs of $I$-band observations to measure the
absolute PMs of individual stars in our target fields. This
is accomplished by determining their shifts with respect to distant
background galaxies. The method we used for this is similar to the
technique that was described and tested extensively in
\citet{soh10,soh12}. We refer the reader to those studies for more
details about the general methodology.

We downloaded the $I$-band {\tt \_flc.fits} images\footnote{The {\tt
\_flc.fits} images are derived from the flat-fielded {\tt \_flt.fits} 
images by application of the \citet{and10} algorithm that corrects 
for imperfect Charge Transfer Efficiency (CTE).} for the first and 
second epochs from the STScI \hst\ archive. We then determined a 
position and a flux for each star in each exposure using the 
{\tt img2xym\_WFC.09x10} program \citep{and06}. The positions were 
corrected for the known ACS/WFC geometric distortions \citep{and06} 
to obtain positions in geometrically rectified frames. Separate
distortion solutions were used for the first- and second-epoch data 
to account for a difference between data taken before and after
\hst\ servicing mission SM4 \citep[see][for details]{soh10,soh12}. 
For each target field, the first exposure of the first epoch (or the
second epoch in the case of field {\bf FIELD 2}, since that has much
deeper data) was selected as the frame of reference. 
We cross-identified all stars in this exposure with the same stars 
in the other exposures. The distortion-corrected positions of the
cross-identified stars were then used to construct a six-parameter
linear transformation between each individual image and the the
reference image.\footnote{The six parameters involve x-y 
translation, scale, rotation, and two components of skew.} 
We then used these transformations to construct a high-resolution 
stacked image for each field, with rejection of cosmic rays and image 
artifacts. For better sampling, the stacked images were super-sampled 
by a factor of two relative to the native ACS/WFC pixel scale.

We identified stars and background galaxies from the stacked image 
for each target field. Stars were selected using the quality-of-fit
parameter reported by the {\tt img2xym\_WFC.09x10} program. For
background galaxies, we started with catalogs generated by running
{\tt SExtractor} \citep{ber96} on the stacked image, and then
inspected each candidate source visually to select only bright 
and compact galaxies.

For each star/galaxy in each of our $I$-band exposures of each target
field, we then measured a position using the template-fitting method
described in \citet{soh10,soh12}. These positions were then corrected
for the geometric distortion as before, for use in our subsequent
analysis.

The templates we used were constructed from the high-resolution
stacked images via a bicubic convolution interpolation method. The
templates were directly fitted to the stars and galaxies in exposures
of the same epoch for which the stacked images were created.  But when
fitting templates to exposures in the other epoch, we applied
additional 7$\times$7 pixel convolution kernels to account for small
PSF differences between the two epochs. For {\bf FIELD 2}, these
kernels were derived using the many bright and isolated stars in the
field (members of the globular cluster NGC 6652), similar to what was
done in our analyses of M31 and Leo~I \citep{soh12,soh13}. However,
for {\bf FIELDS 1}, {\bf 3}, and {\bf 4}, there were not enough stars
in the field to derive reliable kernels. So for these fields we used
an alternative method based on library PSFs, as described in
Appendix~A.

\subsection{Proper Motion Measurements}
\label{sec:starpm}

The next step in the analysis is to transform the
geometrically-corrected, template-fitted positions for all sources in
all exposures into the reference frame. For this we used a different
procedure for {\bf FIELD 2} than for the other fields. {\bf FIELD 2}
is a dense globular cluster field, whereas the other fields are sparse
deep fields with very few stars. We describe the procedure for each of
these situations in turn.

For {\bf FIELD 2}, NGC~6652 stars that are distributed throughout the
frame were used as reference sources. Since the majority of the stars
in this field are NGC~6652 stars, it was easy to select them using
their position in CMD and PM diagrams, as discussed in
Section~\ref{sec:sgrstream} below. The distortion-corrected positions
of these NGC~6652 stars were used to determine six-parameter linear
transformations with respect to the reference frame. These linear
transformations were used to transform the measured positions of {\it
  all} stars and background galaxies in all exposures of both epochs
into the reference frame. In the reference frame, the mean PM of
NGC~6652 is now zero by construction, so that it sets the astrometric
zero point.

The individual exposures yield multiple determinations for the
position of each star or background galaxy in each epoch. We compute
the mean (with outlier rejection) of these determinations to obtain
the average position of each source in each epoch. In addition, the
rms scatter of the multiple measurements in a given epoch yields the
random positional uncertainty in a single measurement, and the error
in the mean is then the rms scatter divided by $\sqrt{N}$, where $N$
is the number of measurements in the epoch. 

To obtain the absolute proper motion $\mu_{i}$ of each star $i$ in
{\bf FIELD 2}, we first measured its relative motion $\mu'_{i}$ with
respect to the bulk motion of NGC~6652. This was obtained by taking
the difference in reference-frame position between the second and the 
first epoch, and dividing by the time baseline. Similarly, we measured
the apparent relative PM $\mu'_{j}$ of each background galaxy $j$
with respect to the bulk motion of NGC~6652. We then took the weighted
average of the $\mu'_{j}$ over all background galaxies to obtain the
bulk motion ${\overline {\mu}'_{\rm bg}}$ of all the background
galaxies with respect to the bulk motion of NGC~6652.

The absolute PM of a given star is then 
\begin{equation}
\label{en:pm_field2}
  \mu_{i} = \mu'_{i} - {\overline {\mu}'_{\rm bg}} .
\end{equation}
Random errors were propagated by adding the errors in each step 
in quadrature, i.e., 
\begin{equation}
\label{eqn:pmerr_field2}
  \Delta {\mu_{i}} = \sqrt{\Delta {\mu'_{i}}^{2} + 
                           \Delta {\overline {\mu}'_{\rm bg}}^{2}} .
\end{equation}
As a byproduct of this process, we also obtain the absolute PM of 
NGC~6652, $\mu_{\rm NGC6652} = - {\overline {\mu}'_{\rm
bg}}$, which will be discussed in Appendix~B.

For {\bf FIELDS 1}, {\bf 3}, and {\bf 4}, the distortion-corrected
positions of background galaxies were directly used to define the
astrometric zero point.\footnote{The number of background galaxies 
that set the zero point of our PMs for each field is listed in
Table~\ref{tab:pmresults}.} Again, for each individual exposure, we
determined six-parameter linear transformations to match the positions
of background galaxies to those in the reference frame. We then used
these transformations to transform the measured positions of {\it all}
sources in all exposures of both epochs into the reference frame.

Since in this approach the frame of reference is defined using distant
galaxies in the background, the absolute PM of a given star is simply
equal to the difference between the average second- and first-epoch
reference frame positions, divided by the time baseline.  As before,
the PM error for each star is given by
equation~(\ref{eqn:pmerr_field2}).

The PMs and their associated errors along the detector axes were
transformed to the directions west and north using the orientation of
each reference image with respect to the sky.

A consistency check on the accuracy of our astrometric reference frame
is provided by the fact that one of our fields contains a known distant
QSO that is unsaturated in our science exposures.  Using the methods
described above, we determined the absolute PM of the QSO in 
{\bf FIELD 3} with respect to the reference frame defined by the 
distant background galaxies. This yields, $(\muw, \mun) = (0.034 \pm 
0.100, 0.068 \pm 0.101)\ \masyr$,\footnote{$\muw$ and $\mun$ are 
defined as the PMs in west ($\muw \equiv -\mu_{\alpha}\cos\delta$) and 
north ($\mun \equiv \mu_{\delta}$) directions, respectively.} which 
is  consistent with zero as expected.

In principle, one could also have used the QSO to set the absolute
reference frame in {\bf FIELD 3}, instead of the background galaxies
\citep[as in, e.g., the study of the Magellanic Clouds by][]{kal13}. 
However, this yields lower accuracy, because the positions of a group 
of background galaxies can on average be determined more accurately 
than the position of a single QSO.

The techniques and data we used here to determine PMs are similar to
those used in our previous work on M31 and Leo~I \citep{soh12,soh13}.
The techniques have many built-in features to reduce the potential for
systematic errors. Based on extensive tests and comparisons performed
in the context of these prior studies \citep[as summarized in
Section~2.4 of][]{soh13}, we expect residual systematic PM errors to
be $\lta 0.03\ \masyr$. This is lower than the random PM measurement
errors that we present below, both for individual stars and for the
overall stellar population in the Sgr stream.  Possible systematic
errors should therefore have a minimal impact on any data-model
comparisons.

\subsection{Photometric Measurements}
\label{sec:photcal}

Photometric measurements of stars in our target fields are
automatically carried out by the {\tt img2xym\_WFC.09x10} program 
when measuring library PSF-based positions in the early stage of data
analysis. These measurements are in instrumental magnitudes (counts)
To calibrate the photometry to the VEGAMAG system, we applied the 
time-dependent zero points provided by STScI.
\footnote{\url{http://www.stsci.edu/hst/acs/analysis/zeropoints}.}
To determine the appropriate aperture corrections, we made use of the
multi-drizzled images of our target fields provided by the \hst\
archive. On these images we measured the brightnesses of several 
bright and isolated stars using aperture radii of 10 pixels 
($0\farcs5$). The sky backgrounds were measured within 20 to 30 
pixels. To correct for the stellar light contributing to the sky 
background level, we used the encircled energy curves listed in 
Table~3 of \citet{sir05}. We then further corrected our photometry to 
an infinite aperture using Table~5 of \citet{sir05}. We then compared 
the magnitudes to the {\tt img2xym\_WFC.09x10} program results for 
the same stars, to determine the appropriate aperture corrections. 
These were then applied to all stars in our photometric catalogs.

\section{Identification of Sagittarius Stream Stars}
\label{sec:sgrstream}

\subsection{Methodology}
\label{sec:identifying}

Our target fields contain stars that belong to the Sgr stream, but
also contain MW stars in the foreground or background. Since our goal
is to characterize the kinematics of the stream, the identification of
Sgr stream stars is crucial. We do this for each field on the basis of
two main sources of information: (1) the CMD constructed using the
photometry obtained from the measurements in
Section~\ref{sec:photcal}; and (2) the absolute PM $(\muw,\mun)$
vector point diagram obtained from the measurements in
Section~\ref{sec:starpm}. To interpret the information, we make
reference to the predictions of stellar population models and 
dynamical models.

\subsubsection{Isochrones}
\label{sec:isochrones}

To identify which stars in the CMD (Figures~\ref{fig:field1}a
--\ref{fig:field4}a) are consistent with belonging to the Sgr 
stream, we generally overlay fiducial isochrones developed by 
the Dartmouth Stellar Evolution Database \citep[DSED,][]{dot08}.
\footnote{See the following URL for details:
  \url{http://stellar.dartmouth.edu/models/index.html}.} 
For this we use two different stellar populations: an old 
metal-poor (OMP) population with (age, [Fe/H], [$\alpha$/Fe]) 
= (13 Gyr, $-1.8$, $+0.2$); and an intermediate metal-rich 
(IMR) population (5 Gyr, $-0.5$, $0.0$). The age and metallicity 
combinations were selected based on the \hst\ CMD study of the 
field around M54 by \citet{sie07}, and the [$\alpha$/Fe] value 
for each metallicity was chosen based on the [Ti/Fe] versus [Fe/H] 
relation of Sgr stream stars studied by \citet{cho10}. Our 
{\bf FIELD 1} is close to (but not overlapping with) the SA 93 
field of \citet{car12}, and our choice of metallicities is 
consistent with the two peaks in their metallicity distributions.

For the reddening applied to the isochrones, we took the E($B-V$)
value estimated from interpolating the reddening maps of
\citet{sch98}. We added 0.02 of extra reddening as suggested by
\citet{sie11}. The absorption values $A_{\rm F814W}$ and $A_{\rm
 F606W}$ were then adopted from Table~6 of \citet{sch11}.

For the distances applied to the isochrones, we started from the
estimates for each field derived in Paper~II. These are based on
interpolation of the stream's known distance variations as function 
of Sgr longitude, as previously measured from ground-based data by 
\citet{bel06,bel14}, \citet{kop12} and \citet{sla13}. To these
distances we then applied small corrections at the $\sim 10$\% level,
to improve the fit to either the observed CMDs or to other
observational constraints modeled in Paper~II.

The exact choices for the stream's stellar population properties,
reddening, or distance are not critical for the present paper. We are
merely trying to determine here which stars in each observed CMD are
plausibly associated with the stream. We are {\it not} trying to
actually fit the CMD, to determine either the stream's stellar
population properties or its distance. Most of our fields are much 
too sparse to make this practical. Such studies are better carried 
out with ground-based data that cover larger fields of view.

\subsubsection{Sagittarius Stream Dynamical Models}
\label{sec:sagmodels}

In Figures~\ref{fig:field1}c--\ref{fig:field4}c,
we show for comparison as fiducial model for each field the PM
predictions of LM10. Their best-fitting $N$-body model for the Sgr
stream includes the PM for each model particle\footnote{See the
  following URL for details:
  \url{http://www.astro.virginia.edu/~srm4n/Sgr/data.html}.}. To
obtain adequate statistics, we consider for each field those model
particles that are within 1.5\degr\ in both right ascension and
declination from the center of the observed \hst\ field. We use
different symbols to indicate whether particles are part of the
leading or trailing arm, and whether they are part of the first or
second wrap of the stream. We also use different colors as in LM10 to
indicate the time at which each particle was stripped. The primary
goal of showing the LM10 model predictions is to aid in the
identification of features or clumps in the observed PM diagrams. The
goal at this stage is {\it not} to perform detailed quantitative
data-model comparisons. Those are presented in Paper~II.

An important goal of the present study is to determine the PMs along
the Sgr stream, so that they can be compared to model predictions. 
For this reason, we are careful {\it not} to bias our selection of 
Sgr stream stars in each field by whether or not they follow the LM10 
PM predictions. After all, the LM10 model does have some known
limitations and it need not be correct. For example, the model does 
not address the observed bifurcation of the stream on the leading side
\citep{bel06}, and it does not fit the observed distance of the 
stream at large angular distances on the trailing side \citep{bel14}.
Nonetheless, the LM10 models does fit many properties of the Sgr 
stream successfully, and it is helpful for the interpretation of the 
PM diagrams to have some sense of what might plausibly be expected.

\subsubsection{Milky Way Dynamical Models}
\label{sec:MWmodels}

To identify which stars in the PM vector point diagram
(Figures~\ref{fig:field1}b--\ref{fig:field4}b) are consistent with
belonging to the Sgr stream, we note that the stream is thin and
dynamically cold \citep{maj04,mon07}.\footnote{The Sgr stream
is dynamically much colder than the MW halo in which it resides,
but it is not as cold as other thinner MW halo streams such as
the Orphan \citep{cas13} and Pal 5 \citep{ode09} streams.}
Therefore, Sgr stars clump in PM space, as is evident also in the
LM10 model predictions shown in
Figures~\ref{fig:field1}c--\ref{fig:field4}c. This is the primary
property that sets Sgr stream stars apart from other MW populations.

Foreground stars in the MW {\it disk} can generally be separated from
Sgr stars both on the basis of their CMD properties and their PM
properties. The most numerous stars in the disk are faint red dwarfs.
Hence, these provide the main MW disk contamination in our fields.
At the apparent magnitudes of interest, these stars are typically
redder than Sgr stars. Moreover, for these low-luminosity stars to
exceed our observational magnitude limit, they tend to be nearby.
Hence, their PM distribution has a large spread of several $\masyr$ 
(in fact, disk stars in our fields often have PMs that place them 
outside the plot ranges in our PM diagrams). This was shown using 
Besancon MW models \citep{rob03} in \citet[][, Figure~3]{dea13}. 
It is also evident in our observations of {\bf FIELD 2} (see
Figure~\ref{fig:field2}), which will be discussed in
Section~\ref{sec:field2} below. This field lies at Galactic
coordinates $(l,b) = (1.9^{\circ},-12.0^{\circ})$, and therefore has a
very large MW disk (and bulge) contamination. Since the MW disk
population is so much more dispersed in PM space than the Sgr
populations of interest, it does not affect any of our subsequent
discussion.

Discriminating Sgr stars from MW {\it halo} stars is also possible,
but not quite as easy. Since halo stars are also part of a
pre-dominantly old population, and are found at similar distances as
Sgr stars, they cannot be separated purely based on their CMD
location. It is therefore important to understand the predicted
location of halo stars in PM space. The MW halo does not possess
significant rotation. Therefore, to lowest order, the mean transverse
velocity of halo stars at any given distance is merely the projection
of the solar reflex velocity. This defines a mean PM {\it direction}
on the plane of the sky. The {\it size} of the mean PM for halo stars
at a given distance is inversely proportional to the
distance. 

The mean PM of halo stars differs from the mean PM of Sgr stars. Since
our Sun is located close to the Galactic Center, the observed PMs of
distant populations reflect primarily the tangential component of the
motion in the Galactocentric rest frame. In this frame, Sgr stars have
a definite sense of rotation while halo stars do not (for reference, a
velocity difference of $200\ \kms$ at $40\ \kpc$ corresponds to 
$\sim 1\ \masyr$ PM difference). The typical distance of MW halo stars 
at the apparent magnitudes of relevance for our \hst\ data is $\sim 
25\ \kpc$ \citep{dea13}. At this distance, a halo velocity dispersion 
of $\sim 100\ \kms$ corresponds to a PM dispersion of $\sim 
0.8\ \masyr$; but the actual PM dispersion of halo stars is larger 
because of the spread in stellar distances. These arguments imply that 
generally the distributions of Sgr and halo stars will have different 
means in PM space, but they may be partially overlapping. The Sgr 
population can be identified because it is the more concentrated of 
the two. Moreover, because our pointings are located on the Sgr stream, 
we expect to find more Sgr stars than halo stars in our fields.

To quantitatively predict the PM distribution of MW halo stars in our
fields, we used the updated version of the Besancon MW models 
\citep{rob14}. We selected from the Monte-Carlo simulated models only
those stars that pass our CMD selection criteria. We then determined
for each component of the predicted PMs the median value and the
68-percent confidence interval. These are shown as a red cross in each
of Figures~\ref{fig:field1}c, \ref{fig:field3}c, and~\ref{fig:field4}c
(as explained above, the location of each cross corresponds roughly to
the reflex motion of the Sun as seen at the median distance of the
halo sample). Roughly half of the MW halo stars in our fields that
pass our CMD selection criteria are expected within the PM area
spanned by these crosses (since, $0.68^2 = 0.46$). This corresponds to
an expectation value of $\sim 5$ stars; the actual number varies by
field as listed in the figure captions. Due to incompleteness at the
faint end (not modeled here), not all of these stars would actually be
expected in our samples. Since the predicted numbers of stars are so
small, we would expect significant Poisson fluctuations. Moreover, the
real MW halo is likely to have more substructure than the smooth
Besancon models, which would cause additional fluctuations. And
finally, the Besancon models do not reflect all of our latest
understanding of the MW halo \citep[e.g., ][]{dea14}. Therefore, the
predictions should be taken as a rough guide only. We ignore the
contribution of the MW halo for {\bf FIELD 2}, located near the Sgr
dSph, since Sgr stars in this field vastly outnumber any possible MW
halo contamination.

\subsection{{\bf FIELD 1}}
\label{sec:field1}

%
\begin{figure*}
\begin{center}
\includegraphics[width=3.20in]{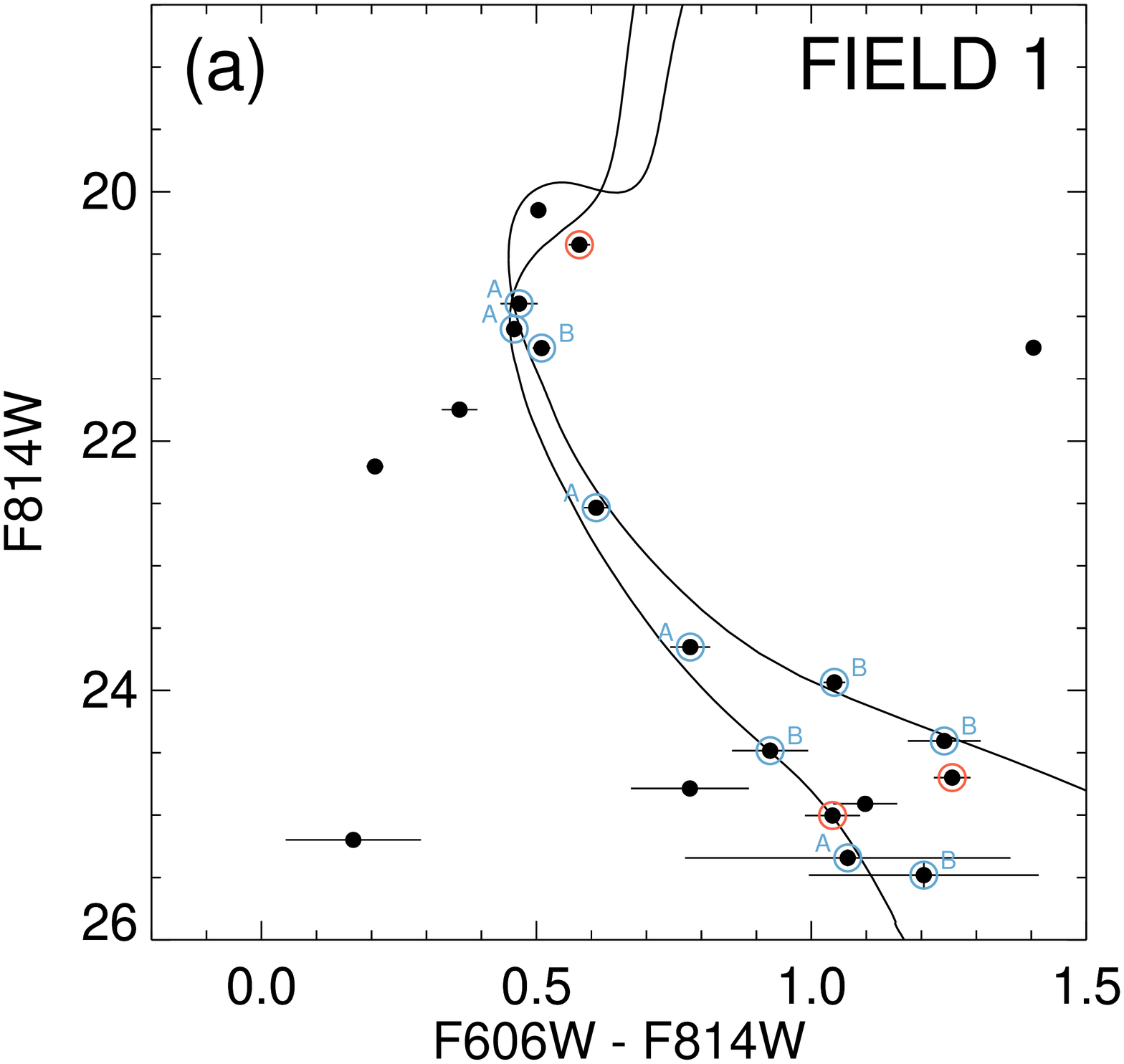}\\
\includegraphics[width=3.20in]{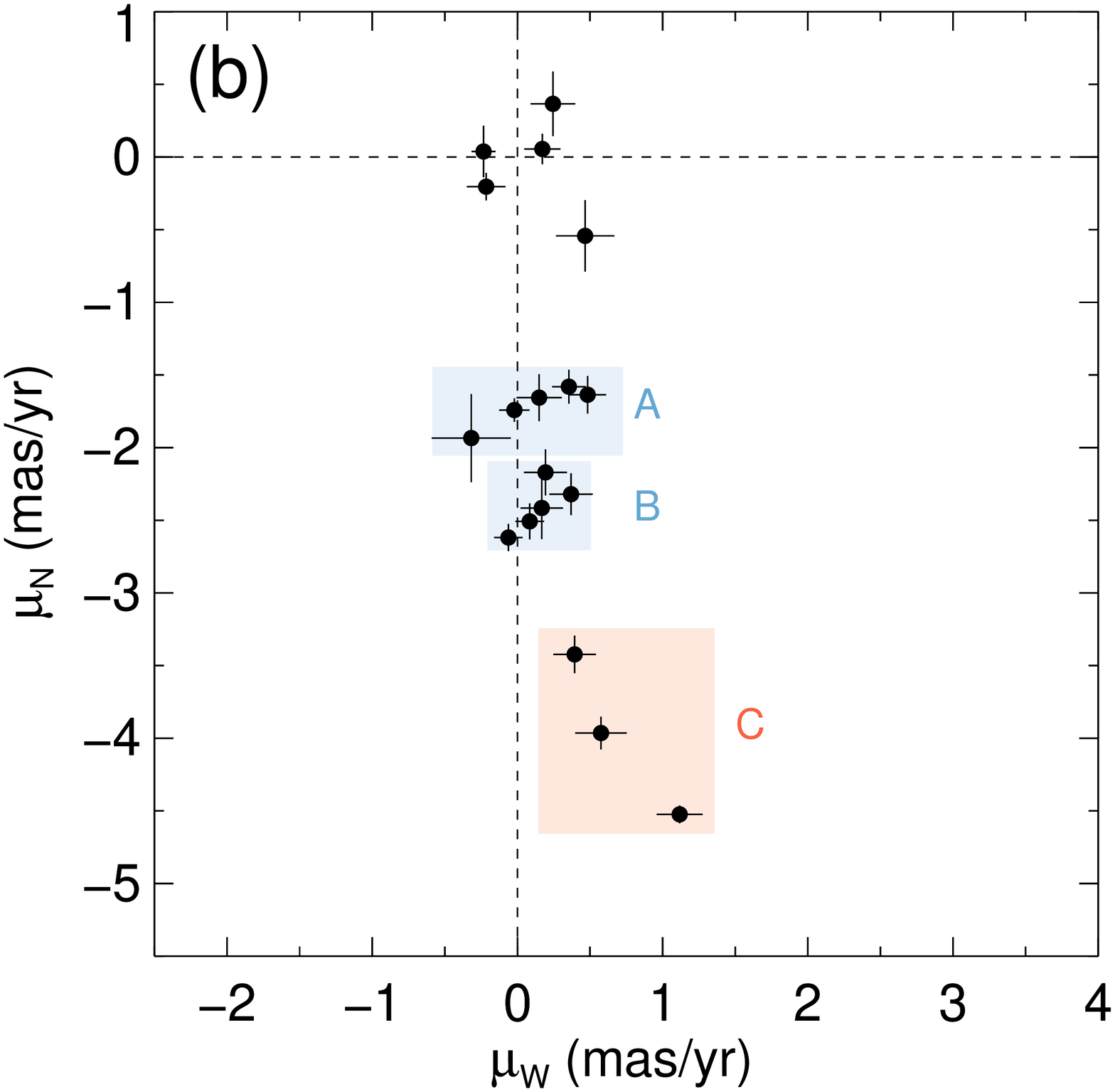}
\includegraphics[width=3.20in]{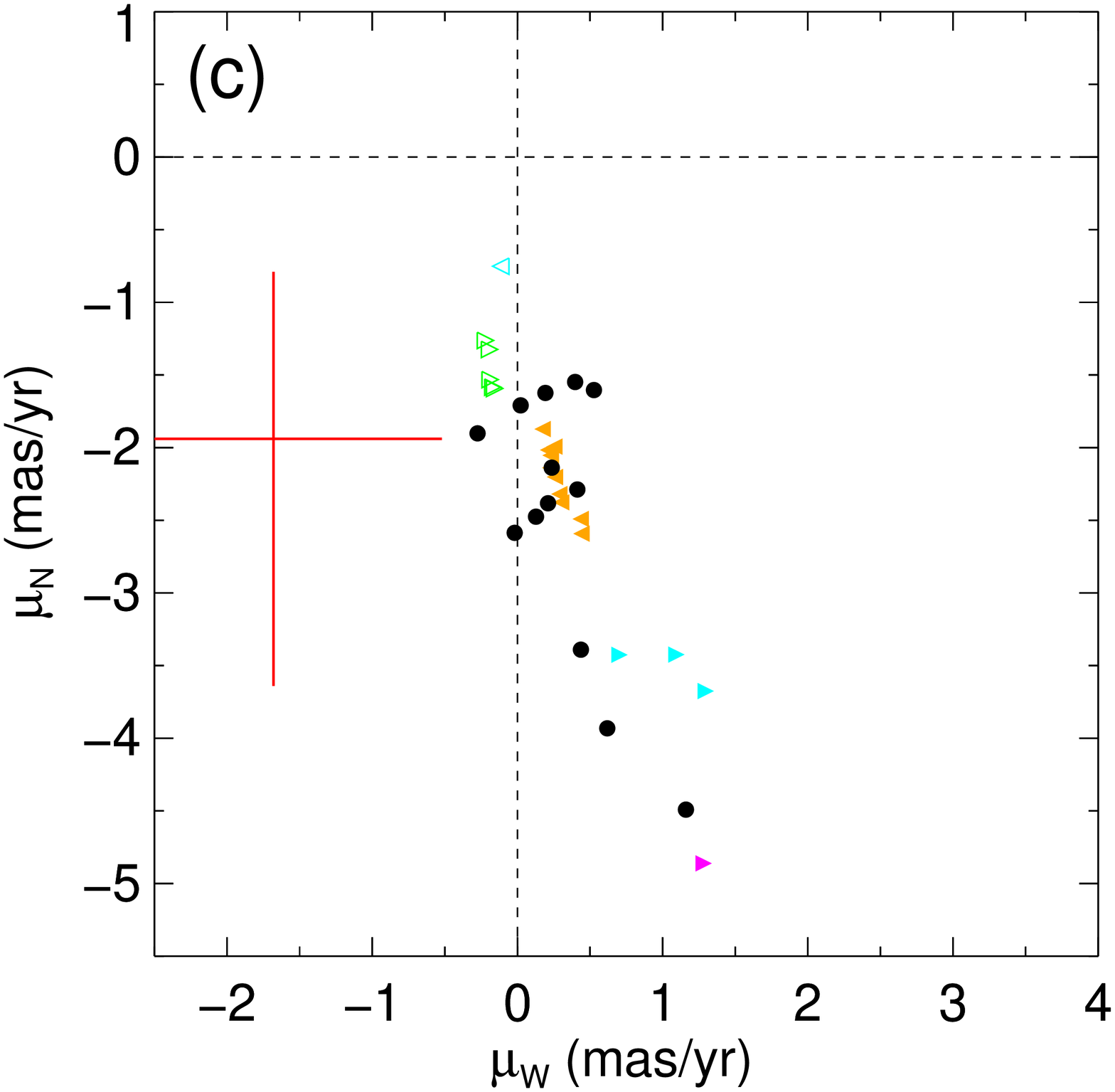}
\end{center}
\caption{Selection of stars associated with the Sgr stream in {\bf
  FIELD 1}. Panel~(a) shows the observed $I$ vs.~$V-I$ CMD;
  panel~(b) shows the observed PM diagram; and panel~(c) shows for
  comparison the PM distribution of $N$-body particles in the model 
  of LM10, as well as the predicted location of MW halo stars (red
  cross).  The observed stars in the field are shown as black symbols
  in all panels, with the associated error bars shown only in panels
  (a) and (b). Not all stars in panel~(a) are shown in panel~(b)
    due to their PMs being outside the range of panel~(b). In
  panel~(c) we only show stars identified as Sgr stream stars, and we
  omit their PM error bars to avoid confusion. The $I$-magnitude
  errors in panel~(a) are smaller than the plot symbols. The black
  curves in panel~(a) are fiducial isochrones for an old metal-poor
  and an intermediate-age metal-rich population (the latter having the
  brighter turn-off and redder colors), as described in the
  text. Colored and labeled boxes in panel~(b) highlight stars that
  likely belong to the Sgr stream, as discussed in the text. The stars
  in these boxes are circled in their corresponding color and labeled
  accordingly in panel~(a). The particles from the LM10 $N$-body model
  in panel~(c) are color-coded following LM10 such that different
  colors represent the time at which a given debris particle became
  unbound from the Sgr dSph: orange for 0 to 1.5 Gyr ago; magenta for
  1.5 to 3 Gyr ago; cyan for 3 to 5 Gyr ago; and green for 5 to 7 Gyr
  ago. Right- and left-facing triangles correspond to stars that
  belong to the leading and trailing arm of Sgr stream, respectively.
  Solid/open triangles correspond to the first/second wrap of the
  stream. The red cross shows the 68 percent confidence intervals in
  each coordinate for MW halo star PMs drawn from Besancon models,
  chosen to meet our CMD selection criteria. An expectation value of 
  3 MW halo stars is predicted within the area spanned by the cross.
\label{fig:field1}}
\end{figure*}
%

Figure~\ref{fig:field1} shows the CMD (panel a), the PM diagram 
(panel b) and the LM10 PM predictions (panel c) for {\bf FIELD 1}. 
This field is located in the trailing arm of the Sgr stream, 
$\sim 104^{\circ}$ from the main body of the Sgr dSph (see
Figure~\ref{fig:fieldlocations}). 
We measured photometry and PMs of all detected stars in this 
field, and rejected stars with 1-D PM errors of $> 0.3\ \masyr$.
A total of 21 stars were considered for further analysis.
Their magnitudes and PMs are shown using black symbols in each panel. 
The adopted distance for the overlaid isochrones in the CMD is 
$33.4\ \kpc$; for comparison, the median distance of the LM10 model 
particles is $30.0\ \kpc$.

Inspection of the CMD shows that most stars fall along one of the two
isochrones. Most of these stars are located along a feature in the PM
diagram that runs roughly from $(\muw, \mun) \approx (0,
-1.5)\ \masyr$ to $(\muw, \mun) \approx (1, -4.5)\ \masyr$.
A similar PM feature is seen in the LM10 model predictions. So even
though the field is sparse, we conclude from the combined CMD and PM
data that we have confidently detected the Sgr stream. The stars
detected in other parts of the PM diagram and with CMD properties that
are inconsistent with the fiducial isochrones are likely foreground or
background objects. 

Based on the apparent clustering in the PM diagram, we detect
four groups of stars. We denote three of the groups as A, B, and C,
as indicated in Figure~\ref{fig:field1}b, and discussed below.
\footnote{Our kinematical groups A, B, and C identified from the 
PM diagrams have no relation to the labeling of the bifurcated 
leading arm (A and B) as shown in Figure~\ref{fig:fieldlocations}.} 
The fourth group consists of five stars near the PM zero point. Most of
these may be unresolved compact background galaxies that are
indistinguishable from stars. This interpretation also seems
reasonable because these objects do not correspond to any kinematical
component predicted by the LM10 model. We note that similar types of
objects were detected in our other PM studies, and in those cases we
drew the same conclusions. For these reasons, we do not discuss this
fourth group of stars further.

None of the groups A, B, or C have the characteristics expected
for a MW halo population (red cross in Figure~\ref{fig:field1}c). 
We therefore identify these groups as Sgr populations. The PMs 
of the individual stars in these groups are listed in
Table~\ref{tab:field1}. It is not obvious that A and B would have to
be physically distinct groups, but the PM diagram does suggest the
possible presence of two separate clumps.  Colored circles and labels
in Figure~\ref{fig:field1}a indicate which star belongs to which
group. There is no unique one-to-one association between the PM groups
and the OMP or IMR isochrones.

Comparison to the LM10 model suggests that stars in group B are
associated with the {\it trailing} arm, and that they have been
recently ($\la 1.5$ Gyr ago) stripped from the main body of the Sgr
dSph. Most likely, the same is true for the stars in group~A. 
Alternatively, their PMs are somewhat consistent with those 
predicted by LM10 for stars in the second wrap of the leading arm. 
But there are as many stars in group~A as group~B, and we would 
normally expect the second wrap to be less populated than the first 
one. So we do not unambiguously detect second-wrap stars in 
{\bf FIELD 1}, but it is possible that they may be present.

The three stars in group C appear to be associated with the {\it
  leading} arm and have been on that part of the stream for at least
3--4 Gyr.  In the LM10 model, these leading-arm stars are a factor
$\sim 2$ closer than the trailing-arm stars. The positions of the C
stars (circled red) in the CMD do not appear inconsistent with this,
but there is not enough information to actually determine their
distance unambiguously.

It is striking that the accuracy of our data is sufficient not only 
to detect the Sgr stream, but even to detect different kinematical
components of the stream. We can identify individual stars {\it in 
the same field} as belonging to either the trailing or the leading 
arm, based on their PMs. This amazing result is made possible due to 
the excellent astrometric capabilities of \hst.

%
\begin{deluxetable}{rcccc}
\tablecaption{Proper motions and photometry of Sgr stream 
stars in {\bf FIELD 1}\label{tab:field1}}
\tablehead{
 \colhead{}      & \colhead{$\muw$}     & \colhead{$\mun$}     & \colhead{F814W}     & \colhead{F606W}      \\
 \colhead{ID}    & \colhead{($\masyr$)} & \colhead{($\masyr$)} & \colhead{(VEGAMAG)} & \colhead{(VEGAMAG)} 
   }
\startdata
\cutinhead{Group A}
  1 &       $-$0.02 $\pm$ 0.10 & $-$1.74 $\pm$ 0.08 & 20.90 $\pm$ 0.01 & 21.37 $\pm$ 0.03 \\
  2 & \phm{$-$}0.35 $\pm$ 0.12 & $-$1.58 $\pm$ 0.12 & 21.10 $\pm$ 0.01 & 21.56 $\pm$ 0.01 \\
  3 & \phm{$-$}0.48 $\pm$ 0.13 & $-$1.64 $\pm$ 0.13 & 22.53 $\pm$ 0.01 & 23.14 $\pm$ 0.02 \\
  4 & \phm{$-$}0.15 $\pm$ 0.16 & $-$1.66 $\pm$ 0.16 & 23.65 $\pm$ 0.02 & 24.43 $\pm$ 0.03 \\
  5 &       $-$0.32 $\pm$ 0.27 & $-$1.93 $\pm$ 0.30 & 25.34 $\pm$ 0.07 & 26.41 $\pm$ 0.29 \\
\cutinhead{Group B}
  6 &       $-$0.06 $\pm$ 0.10 & $-$2.62 $\pm$ 0.09 & 21.25 $\pm$ 0.01 & 21.76 $\pm$ 0.01 \\
  7 & \phm{$-$}0.08 $\pm$ 0.10 & $-$2.51 $\pm$ 0.12 & 23.94 $\pm$ 0.02 & 24.98 $\pm$ 0.00 \\
  8 & \phm{$-$}0.19 $\pm$ 0.15 & $-$2.17 $\pm$ 0.16 & 24.41 $\pm$ 0.03 & 25.65 $\pm$ 0.06 \\
  9 & \phm{$-$}0.37 $\pm$ 0.15 & $-$2.32 $\pm$ 0.14 & 24.48 $\pm$ 0.06 & 25.41 $\pm$ 0.03 \\
 10 & \phm{$-$}0.17 $\pm$ 0.15 & $-$2.42 $\pm$ 0.21 & 25.48 $\pm$ 0.10 & 26.69 $\pm$ 0.18 \\
\cutinhead{Group C}
 11 & \phm{$-$}1.12 $\pm$ 0.16 & $-$4.52 $\pm$ 0.06 & 20.43 $\pm$ 0.01 & 21.00 $\pm$ 0.02 \\
 12 & \phm{$-$}0.58 $\pm$ 0.18 & $-$3.96 $\pm$ 0.11 & 24.70 $\pm$ 0.02 & 25.96 $\pm$ 0.02 \\
 13 & \phm{$-$}0.39 $\pm$ 0.15 & $-$3.42 $\pm$ 0.13 & 25.00 $\pm$ 0.05 & 26.04 $\pm$ 0.02
\enddata
\end{deluxetable}
%

\subsection{{\bf FIELD 2}}
\label{sec:field2}

%
\begin{figure*}
\begin{center}
\includegraphics[width=3.20in]{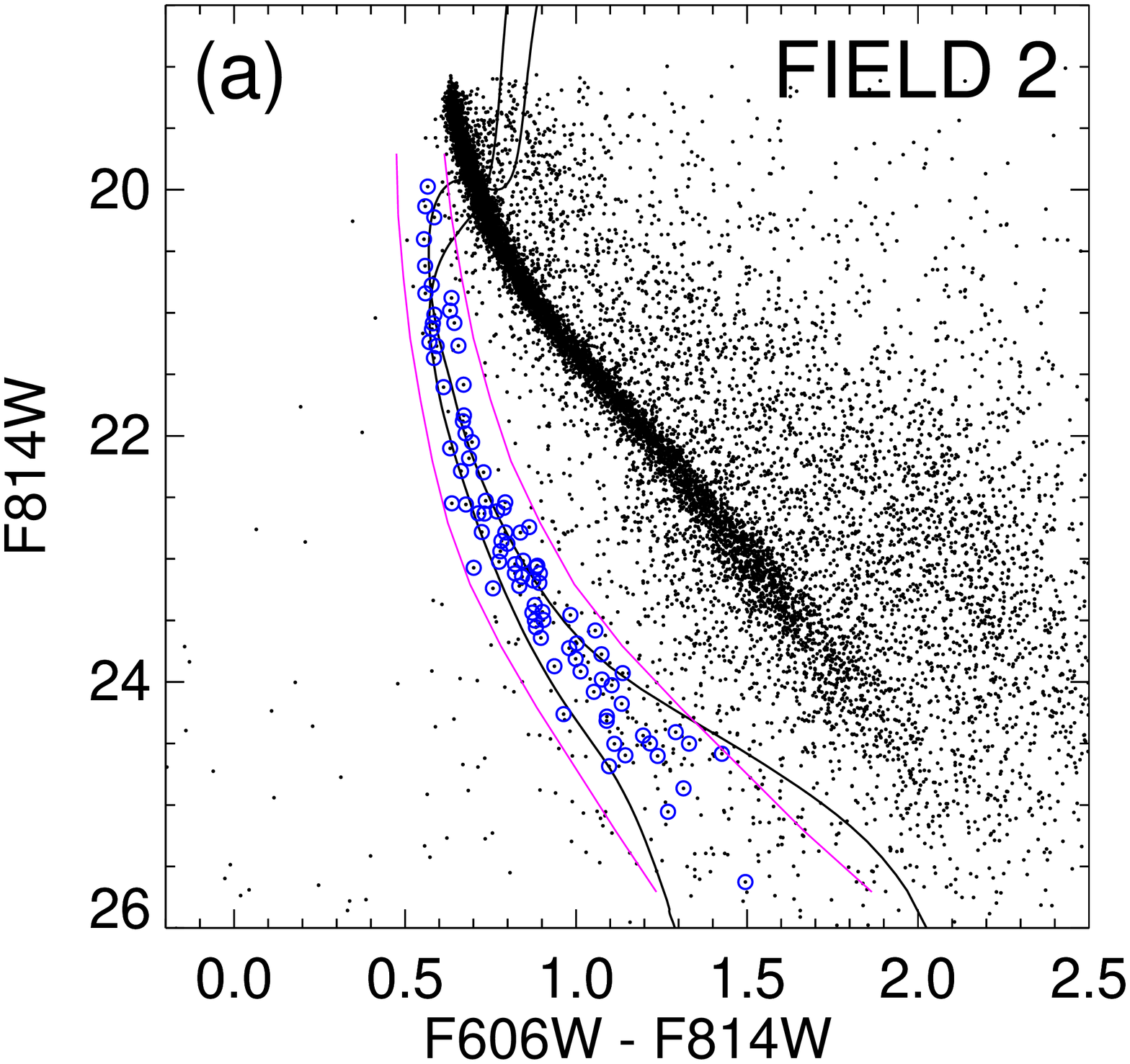}\\
\includegraphics[width=3.20in]{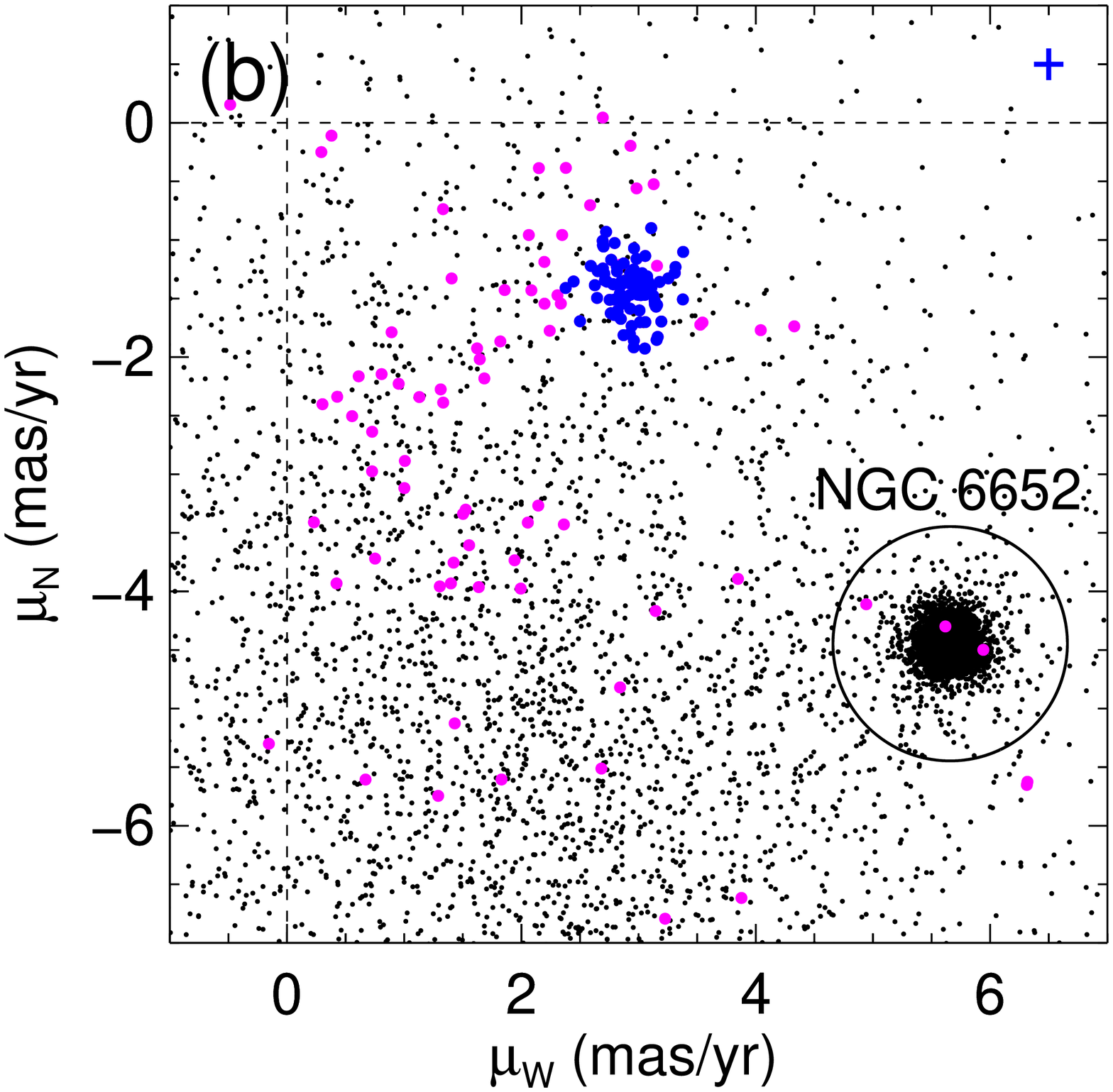}
\includegraphics[width=3.20in]{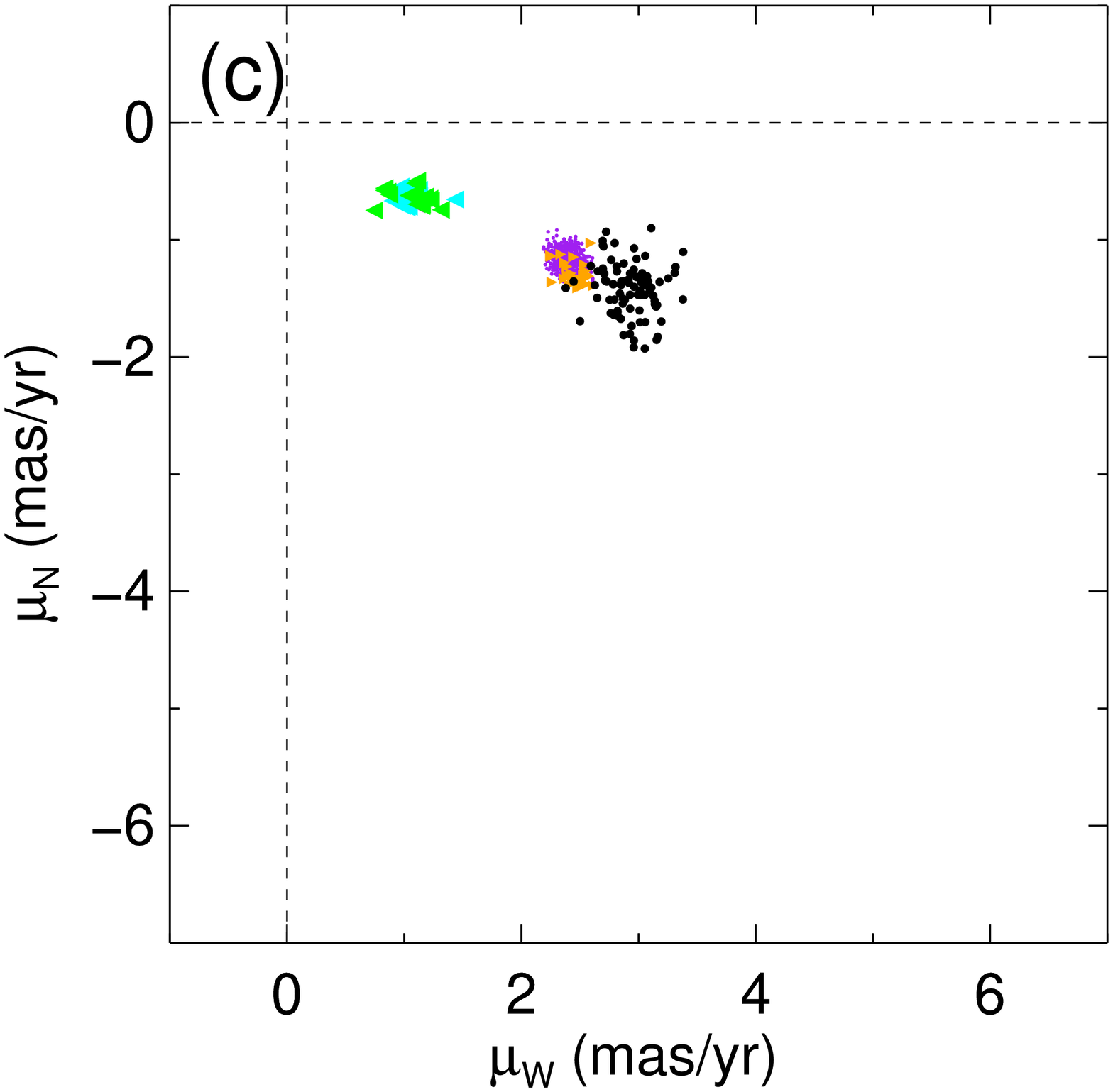}
\end{center}
\caption{Selection of stars associated with the Sgr stream in {\bf
  FIELD 2}. The panels and symbols are generally similar to those in
  Figure~\ref{fig:field1}, but with some exceptions as noted
  below. {\bf FIELD 2} is centered on the globular cluster NGC~6652.
  The globular cluster stars are easily recognized by their MS in the
  CMD, and their clustering in PM space, as indicated in panel (b).
  The Sgr stream stars in the background outline a secondary MS in the
  CMD, blueward of the NGC~6652 MS. We define candidate Sgr stream
  stars as lying between the magenta lines. In panel (b), blue-colored
  stars (also circled blue in panel (a), and shown in black in panel
  (c)) are those that survive a 3-$\sigma$ clipping algorithm in PM
  space. Magenta-colored stars are those that are not consistent at
  the 3-$\sigma$ level with the weighted mean PM of the blue-colored
  stars. Individual PM error bars are not shown to avoid
  confusion. The blue cross in the top right of panel (b) indicates
  the median PM error bars for the blue points. The black points that
  are widely scattered over the PM diagram are mostly Milky Way disk
  and bulge field stars. The groups of green and cyan symbols 
  in panel (c) consists entirely of trailing particles, while the 
  group of orange symbols consists entirely of leading particles.
  Purple dots that largely overlap with the orange triangles are 
  LM10 model particles that are bound to the Sgr dSph. There are 465 
  purple dots and 28 orange triangles near $(\muw, \mun) = 
  (2.4, -1.2)\ \masyr$.
  \label{fig:field2}}
\end{figure*}
%

Figure~\ref{fig:field2} shows the CMD, the PM diagram, and
the LM10 PM predictions for {\bf FIELD 2}. This field is located in
the leading arm of the Sgr stream, but it is close to ($\sim
4^{\circ}$ away from) the main body of the Sgr dSph (see
Figure~\ref{fig:fieldlocations}). The field is centered on the
globular cluster NGC~6652, and its main sequence (MS) dominates the
CMD. But as pointed out by \citet{sie11}, the MS of the Sgr stars in
the background is readily visible as a faint secondary sequence
roughly parallel to and below the MS of NGC~6652. The adopted distance
for the overlaid isochrones in the CMD is $30.5\ \kpc$; for comparison, 
the median distance of the LM10 model particles is $28.4\ \kpc$. 
The distance of NGC~6652 is $10.5\ \kpc$.

Since the angular distance to the Sgr dSph core is much smaller than 
for the other \hst\ fields (see Figure~\ref{fig:fieldlocations}), it 
has a much higher density of Sgr stars. This makes the selection of 
these stars in the CMD straightforward. We define candidate Sgr stars 
as those stars lying between the magenta lines in 
Figure~\ref{fig:field2}a. This yields 90 candidate Sgr stars, out of
$> 12,000$ total stars identified in the field.

The PM distribution of the stars in {\bf FIELD 2}, shown in
Figure~\ref{fig:field2}b, has the two conspicuous clumps that are
readily identified. One clump at $(\muw, \mun) \approx (5.5, -4.5)\ 
\masyr$ is due to stars in NGC 6652. The other clump at $(\muw, \mun)
\approx (2.9, -1.4)\ \masyr$ is due to Sgr stream stars. The remaining
stars that are widely scattered over the PM diagram are mostly MW disk
and bulge field stars. All these results are qualitatively similar to
those reported by \citet{mas13} for NGC 6681, another MW globular
cluster towards the Galactic Bulge that has Sgr stars in the
background.

To obtain a final selection of Sgr stream stars, we first determined
the mean PM of all the candidate stars identified from the CMD. We
then applied an iterative 3-$\sigma$ rejection to reject those stars
(shown in magenta color) with PMs inconsistent with this mean. We
identify the remaining 90 stars (shown in blue) as bona-fide Sgr
stream stars. The PMs of these individual stars are listed in
Table~\ref{tab:field2}.

The observed PM clump in Figure~\ref{fig:field2}c shows 
reasonable agreement with the LM10 $N$-body model particles on 
the leading arm that are either currently bound to the Sgr dSph 
(purple dots) or recently became unbound (orange triangles). 
This indicates that the majority of stars in the 
observed clump are bound to the Sgr dSph and belong to the 
leading arm (as expected, since the field is on the leading arm 
of the Sgr stream). Interestingly, while the PM direction of 
the observed stars is the same as in the model, the size of the 
observed PMs appears to be somewhat larger by $\sim 0.5\ \masyr$.  
We discuss this discrepancy in Section~\ref{sec:PMmodel}.

In this field, the LM10 model also predicts, at a different location
in PM space, the presence of trailing-arm particles (green and cyan 
triangles) that are almost $360^{\circ}$ wrapped from the main body. 
Although this clump of particles appear to be conspicuous in 
Figure~\ref{fig:field2}c, it consist of only 41 model particles,
i.e., $\sim 8$\% of total number of leading-arm model particles.
This means the model predicts about 7 trailing-arm stars in this 
field. We do not find stars with PMs consistent with these particles 
in our data. Given that the trailing-arm model particles lie at  
further distance than the leading-arm particles (at an average 
distance of $\sim 47\ \kpc$), we expanded our search in the CMD 
to look for the trailing-arm stars but did not find any. This may 
indicate that the model overpredicts the density of particles at 
large angular distances from the main body of the Sgr dSph, or that 
it incorrectly predicts their location. Alternatively, it may simply 
be due to small number statistics.

%
\begin{deluxetable}{rcccc}
\tablecaption{Proper motions and photometry of Sgr stream 
stars in {\bf FIELD 2}\label{tab:field2}}
\tablehead{
 \colhead{}      & \colhead{$\muw$}     & \colhead{$\mun$}     & \colhead{F814W}     & \colhead{F606W}      \\
 \colhead{ID}    & \colhead{($\masyr$)} & \colhead{($\masyr$)} & \colhead{(VEGAMAG)} & \colhead{(VEGAMAG)} 
   }
\startdata
  1 & 3.04 $\pm$ 0.10 & $-$1.39 $\pm$ 0.22 & 19.97 $\pm$ 0.00 & 20.54 $\pm$ 0.01 \\
  2 & 2.90 $\pm$ 0.11 & $-$1.38 $\pm$ 0.13 & 20.13 $\pm$ 0.01 & 20.69 $\pm$ 0.01 \\
  3 & 3.02 $\pm$ 0.09 & $-$1.36 $\pm$ 0.11 & 20.22 $\pm$ 0.01 & 20.81 $\pm$ 0.00 \\
  4 & 3.00 $\pm$ 0.09 & $-$1.37 $\pm$ 0.13 & 20.40 $\pm$ 0.01 & 20.96 $\pm$ 0.10 \\
  5 & 3.09 $\pm$ 0.09 & $-$1.42 $\pm$ 0.14 & 20.62 $\pm$ 0.01 & 21.18 $\pm$ 0.01 \\
  6 & 2.86 $\pm$ 0.08 & $-$1.37 $\pm$ 0.11 & 20.77 $\pm$ 0.01 & 21.35 $\pm$ 0.01 \\
  7 & 3.13 $\pm$ 0.16 & $-$1.58 $\pm$ 0.14 & 20.84 $\pm$ 0.00 & 21.40 $\pm$ 0.01 \\
  8 & 3.05 $\pm$ 0.11 & $-$1.32 $\pm$ 0.11 & 20.88 $\pm$ 0.01 & 21.52 $\pm$ 0.00 \\
  9 & 2.94 $\pm$ 0.10 & $-$1.09 $\pm$ 0.14 & 20.98 $\pm$ 0.05 & 21.61 $\pm$ 0.01 \\
 10 & 2.68 $\pm$ 0.21 & $-$1.26 $\pm$ 0.14 & 21.02 $\pm$ 0.01 & 21.60 $\pm$ 0.00 \\
 11 & 2.86 $\pm$ 0.09 & $-$1.21 $\pm$ 0.18 & 21.08 $\pm$ 0.01 & 21.73 $\pm$ 0.00 \\
 12 & 2.80 $\pm$ 0.11 & $-$1.62 $\pm$ 0.14 & 21.09 $\pm$ 0.01 & 21.67 $\pm$ 0.01 \\
 13 & 2.99 $\pm$ 0.11 & $-$1.45 $\pm$ 0.16 & 21.13 $\pm$ 0.01 & 21.71 $\pm$ 0.02 \\
 14 & 2.75 $\pm$ 0.15 & $-$1.18 $\pm$ 0.19 & 21.24 $\pm$ 0.01 & 21.81 $\pm$ 0.01 \\
 15 & 3.24 $\pm$ 0.10 & $-$1.34 $\pm$ 0.11 & 21.27 $\pm$ 0.01 & 21.92 $\pm$ 0.01 \\
 16 & 2.76 $\pm$ 0.11 & $-$1.39 $\pm$ 0.15 & 21.27 $\pm$ 0.01 & 21.86 $\pm$ 0.01 \\
 17 & 3.01 $\pm$ 0.09 & $-$1.38 $\pm$ 0.14 & 21.37 $\pm$ 0.01 & 21.95 $\pm$ 0.06 \\
 18 & 2.78 $\pm$ 0.16 & $-$1.52 $\pm$ 0.12 & 21.58 $\pm$ 0.01 & 22.26 $\pm$ 0.02 \\
 19 & 2.70 $\pm$ 0.12 & $-$1.36 $\pm$ 0.12 & 21.60 $\pm$ 0.01 & 22.22 $\pm$ 0.01 \\
 20 & 3.03 $\pm$ 0.13 & $-$1.40 $\pm$ 0.14 & 21.83 $\pm$ 0.02 & 22.51 $\pm$ 0.01 \\
 21 & 2.94 $\pm$ 0.08 & $-$1.27 $\pm$ 0.22 & 21.88 $\pm$ 0.01 & 22.55 $\pm$ 0.01 \\
 22 & 2.90 $\pm$ 0.22 & $-$1.35 $\pm$ 0.17 & 21.98 $\pm$ 0.01 & 22.66 $\pm$ 0.00 \\
 23 & 2.92 $\pm$ 0.15 & $-$1.75 $\pm$ 0.17 & 22.05 $\pm$ 0.01 & 22.75 $\pm$ 0.01 \\
 24 & 2.68 $\pm$ 0.11 & $-$1.06 $\pm$ 0.15 & 22.10 $\pm$ 0.01 & 22.73 $\pm$ 0.00 \\
 25 & 3.08 $\pm$ 0.11 & $-$1.44 $\pm$ 0.12 & 22.18 $\pm$ 0.01 & 22.87 $\pm$ 0.00 \\
 26 & 2.85 $\pm$ 0.14 & $-$1.52 $\pm$ 0.18 & 22.29 $\pm$ 0.01 & 22.95 $\pm$ 0.01 \\
 27 & 2.94 $\pm$ 0.13 & $-$1.42 $\pm$ 0.19 & 22.30 $\pm$ 0.01 & 23.03 $\pm$ 0.02 \\
 28 & 2.32 $\pm$ 0.15 & $-$1.56 $\pm$ 0.19 & 22.53 $\pm$ 0.02 & 23.26 $\pm$ 0.00 \\
 29 & 2.63 $\pm$ 0.11 & $-$1.51 $\pm$ 0.17 & 22.54 $\pm$ 0.01 & 23.33 $\pm$ 0.01 \\
 30 & 3.16 $\pm$ 0.19 & $-$1.37 $\pm$ 0.17 & 22.55 $\pm$ 0.02 & 23.19 $\pm$ 0.14 \\
 31 & 2.61 $\pm$ 0.10 & $-$1.40 $\pm$ 0.13 & 22.56 $\pm$ 0.02 & 23.24 $\pm$ 0.02 \\
 32 & 3.01 $\pm$ 0.10 & $-$1.30 $\pm$ 0.13 & 22.59 $\pm$ 0.03 & 23.37 $\pm$ 0.01 \\
 33 & 2.83 $\pm$ 0.15 & $-$1.40 $\pm$ 0.14 & 22.61 $\pm$ 0.02 & 23.38 $\pm$ 0.02 \\
 34 & 2.69 $\pm$ 0.14 & $-$1.30 $\pm$ 0.17 & 22.63 $\pm$ 0.02 & 23.34 $\pm$ 0.02 \\
 35 & 3.04 $\pm$ 0.21 & $-$1.48 $\pm$ 0.14 & 22.63 $\pm$ 0.01 & 23.36 $\pm$ 0.02 \\
 36 & 3.13 $\pm$ 0.12 & $-$1.58 $\pm$ 0.21 & 22.74 $\pm$ 0.00 & 23.60 $\pm$ 0.03 \\
 37 & 2.91 $\pm$ 0.11 & $-$1.48 $\pm$ 0.16 & 22.78 $\pm$ 0.02 & 23.51 $\pm$ 0.03 \\
 38 & 3.11 $\pm$ 0.10 & $-$1.49 $\pm$ 0.21 & 22.79 $\pm$ 0.03 & 23.58 $\pm$ 0.03 \\
 39 & 2.96 $\pm$ 0.10 & $-$1.18 $\pm$ 0.15 & 22.79 $\pm$ 0.02 & 23.62 $\pm$ 0.05 \\
 40 & 3.07 $\pm$ 0.16 & $-$1.42 $\pm$ 0.15 & 22.85 $\pm$ 0.03 & 23.64 $\pm$ 0.02 \\
 41 & 2.71 $\pm$ 0.15 & $-$1.37 $\pm$ 0.14 & 22.88 $\pm$ 0.02 & 23.68 $\pm$ 0.01 \\
 42 & 3.00 $\pm$ 0.16 & $-$1.48 $\pm$ 0.21 & 22.94 $\pm$ 0.01 & 23.72 $\pm$ 0.02 \\
 43 & 3.15 $\pm$ 0.19 & $-$1.84 $\pm$ 0.24 & 23.02 $\pm$ 0.03 & 23.86 $\pm$ 0.27 \\
 44 & 2.99 $\pm$ 0.20 & $-$1.72 $\pm$ 0.26 & 23.02 $\pm$ 0.02 & 23.80 $\pm$ 0.03 \\
 45 & 2.43 $\pm$ 0.08 & $-$1.37 $\pm$ 0.11 & 23.04 $\pm$ 0.04 & 23.86 $\pm$ 0.00 \\
 46 & 2.36 $\pm$ 0.15 & $-$1.42 $\pm$ 0.17 & 23.06 $\pm$ 0.06 & 23.94 $\pm$ 0.29 \\
 47 & 2.83 $\pm$ 0.17 & $-$1.69 $\pm$ 0.23 & 23.07 $\pm$ 0.02 & 23.95 $\pm$ 0.03 \\
 48 & 2.77 $\pm$ 0.13 & $-$1.39 $\pm$ 0.16 & 23.07 $\pm$ 0.01 & 23.77 $\pm$ 0.15 \\
 49 & 2.94 $\pm$ 0.11 & $-$1.87 $\pm$ 0.16 & 23.12 $\pm$ 0.02 & 23.94 $\pm$ 0.01 \\
 50 & 3.12 $\pm$ 0.11 & $-$1.56 $\pm$ 0.14 & 23.12 $\pm$ 0.04 & 24.01 $\pm$ 0.01 \\
 51 & 2.82 $\pm$ 0.18 & $-$1.23 $\pm$ 0.21 & 23.15 $\pm$ 0.03 & 23.99 $\pm$ 0.00 \\
 52 & 2.80 $\pm$ 0.13 & $-$1.28 $\pm$ 0.22 & 23.18 $\pm$ 0.02 & 24.05 $\pm$ 0.05 \\
 53 & 2.80 $\pm$ 0.12 & $-$1.24 $\pm$ 0.13 & 23.20 $\pm$ 0.04 & 24.09 $\pm$ 0.01 \\
 54 & 2.91 $\pm$ 0.19 & $-$1.30 $\pm$ 0.22 & 23.22 $\pm$ 0.02 & 24.05 $\pm$ 0.03 \\
 55 & 2.82 $\pm$ 0.24 & $-$1.47 $\pm$ 0.17 & 23.24 $\pm$ 0.02 & 24.00 $\pm$ 0.14 \\
 56 & 2.68 $\pm$ 0.31 & $-$1.07 $\pm$ 0.20 & 23.38 $\pm$ 0.03 & 24.25 $\pm$ 0.04 \\
 57 & 2.78 $\pm$ 0.15 & $-$1.04 $\pm$ 0.24 & 23.43 $\pm$ 0.02 & 24.33 $\pm$ 0.05 \\
 58 & 2.63 $\pm$ 0.22 & $-$1.28 $\pm$ 0.22 & 23.44 $\pm$ 0.04 & 24.31 $\pm$ 0.03 \\
 59 & 3.14 $\pm$ 0.27 & $-$1.87 $\pm$ 0.21 & 23.46 $\pm$ 0.08 & 24.44 $\pm$ 0.05 \\
 60 & 2.57 $\pm$ 0.15 & $-$1.24 $\pm$ 0.22 & 23.50 $\pm$ 0.02 & 24.40 $\pm$ 0.02 \\
 61 & 2.18 $\pm$ 0.16 & $-$1.20 $\pm$ 0.17 & 23.50 $\pm$ 0.02 & 24.38 $\pm$ 0.03 \\
 62 & 2.86 $\pm$ 0.14 & $-$1.53 $\pm$ 0.11 & 23.56 $\pm$ 0.03 & 24.44 $\pm$ 0.12 \\
 63 & 2.83 $\pm$ 0.14 & $-$1.37 $\pm$ 0.17 & 23.58 $\pm$ 0.02 & 24.64 $\pm$ 0.03 \\
 64 & 3.14 $\pm$ 0.14 & $-$1.23 $\pm$ 0.18 & 23.64 $\pm$ 0.03 & 24.54 $\pm$ 0.01 \\
 65 & 2.85 $\pm$ 0.23 & $-$1.56 $\pm$ 0.23 & 23.69 $\pm$ 0.03 & 24.69 $\pm$ 0.03 \\
 66 & 2.80 $\pm$ 0.16 & $-$1.63 $\pm$ 0.17 & 23.73 $\pm$ 0.04 & 24.71 $\pm$ 0.02 \\
 67 & 2.74 $\pm$ 0.20 & $-$1.53 $\pm$ 0.20 & 23.78 $\pm$ 0.05 & 24.85 $\pm$ 0.07 \\
 68 & 3.09 $\pm$ 0.26 & $-$0.91 $\pm$ 0.35 & 23.81 $\pm$ 0.06 & 24.81 $\pm$ 0.02 \\
 69 & 2.48 $\pm$ 0.21 & $-$1.71 $\pm$ 0.19 & 23.87 $\pm$ 0.03 & 24.81 $\pm$ 0.03 \\
 70 & 3.12 $\pm$ 0.15 & $-$1.53 $\pm$ 0.48 & 23.91 $\pm$ 0.04 & 24.93 $\pm$ 0.05
\enddata
\end{deluxetable}

\setcounter{table}{2}
\begin{deluxetable}{rcccc}
\tablecaption{Continued from previous page.}
\tablehead{
 \colhead{}      & \colhead{$\muw$}     & \colhead{$\mun$}     & \colhead{F814W}     & \colhead{F606W}      \\
 \colhead{ID}    & \colhead{($\masyr$)} & \colhead{($\masyr$)} & \colhead{(VEGAMAG)} & \colhead{(VEGAMAG)} 
   }
\startdata
 71 & 3.04 $\pm$ 0.18 & $-$1.15 $\pm$ 0.21 & 23.93 $\pm$ 0.07 & 25.06 $\pm$ 0.03 \\
 72 & 2.77 $\pm$ 0.33 & $-$1.65 $\pm$ 0.22 & 23.98 $\pm$ 0.03 & 25.06 $\pm$ 0.05 \\
 73 & 2.99 $\pm$ 0.26 & $-$1.62 $\pm$ 0.30 & 24.03 $\pm$ 0.06 & 25.13 $\pm$ 0.02 \\
 74 & 3.29 $\pm$ 0.18 & $-$1.30 $\pm$ 0.18 & 24.08 $\pm$ 0.04 & 25.13 $\pm$ 0.07 \\
 75 & 3.14 $\pm$ 0.29 & $-$1.57 $\pm$ 0.21 & 24.18 $\pm$ 0.03 & 25.31 $\pm$ 0.01 \\
 76 & 2.96 $\pm$ 0.20 & $-$1.33 $\pm$ 0.44 & 24.26 $\pm$ 0.06 & 25.23 $\pm$ 0.06 \\
 77 & 2.80 $\pm$ 0.25 & $-$1.66 $\pm$ 0.42 & 24.28 $\pm$ 0.04 & 25.37 $\pm$ 0.01 \\
 78 & 2.91 $\pm$ 0.35 & $-$1.60 $\pm$ 0.36 & 24.32 $\pm$ 0.04 & 25.41 $\pm$ 0.06 \\
 79 & 2.85 $\pm$ 0.24 & $-$1.83 $\pm$ 0.27 & 24.41 $\pm$ 0.06 & 25.70 $\pm$ 0.10 \\
 80 & 2.74 $\pm$ 0.35 & $-$1.64 $\pm$ 0.40 & 24.44 $\pm$ 0.05 & 25.63 $\pm$ 0.02 \\
 81 & 2.70 $\pm$ 0.22 & $-$0.94 $\pm$ 0.20 & 24.50 $\pm$ 0.04 & 25.72 $\pm$ 0.05 \\
 82 & 3.06 $\pm$ 0.28 & $-$1.37 $\pm$ 0.49 & 24.50 $\pm$ 0.07 & 25.83 $\pm$ 0.16 \\
 83 & 3.30 $\pm$ 0.45 & $-$1.24 $\pm$ 0.34 & 24.50 $\pm$ 0.07 & 25.62 $\pm$ 0.08 \\
 84 & 3.36 $\pm$ 0.42 & $-$1.12 $\pm$ 0.49 & 24.58 $\pm$ 0.04 & 26.01 $\pm$ 0.03 \\
 85 & 2.68 $\pm$ 0.39 & $-$1.02 $\pm$ 0.31 & 24.60 $\pm$ 0.04 & 25.74 $\pm$ 0.04 \\
 86 & 3.18 $\pm$ 0.40 & $-$1.71 $\pm$ 0.65 & 24.60 $\pm$ 0.07 & 25.84 $\pm$ 0.05 \\
 87 & 3.51 $\pm$ 0.44 & $-$1.74 $\pm$ 0.30 & 24.69 $\pm$ 0.06 & 25.78 $\pm$ 0.02 \\
 88 & 3.04 $\pm$ 0.56 & $-$1.94 $\pm$ 0.61 & 24.86 $\pm$ 0.05 & 26.18 $\pm$ 0.22 \\
 89 & 2.18 $\pm$ 0.25 & $-$1.56 $\pm$ 0.38 & 25.06 $\pm$ 0.08 & 26.32 $\pm$ 0.13 \\
 90 & 2.29 $\pm$ 0.39 & $-$1.49 $\pm$ 0.33 & 25.63 $\pm$ 0.11 & 27.12 $\pm$ 0.22
\enddata
\end{deluxetable}
%

\subsection{{\bf FIELD 3}}
\label{sec:field3}

%
\begin{figure*}
\begin{center}
\includegraphics[width=3.20in]{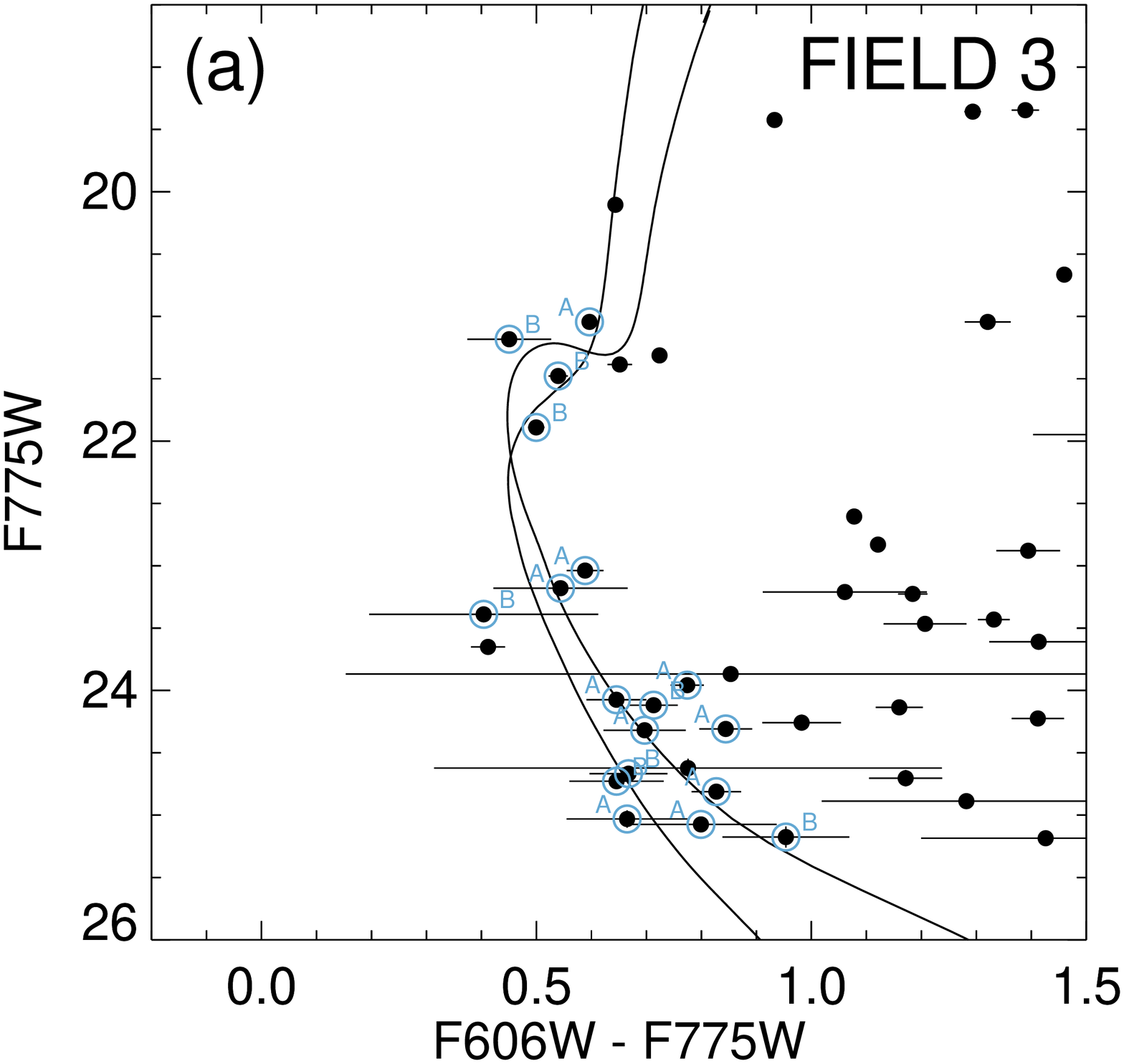}\\
\includegraphics[width=3.20in]{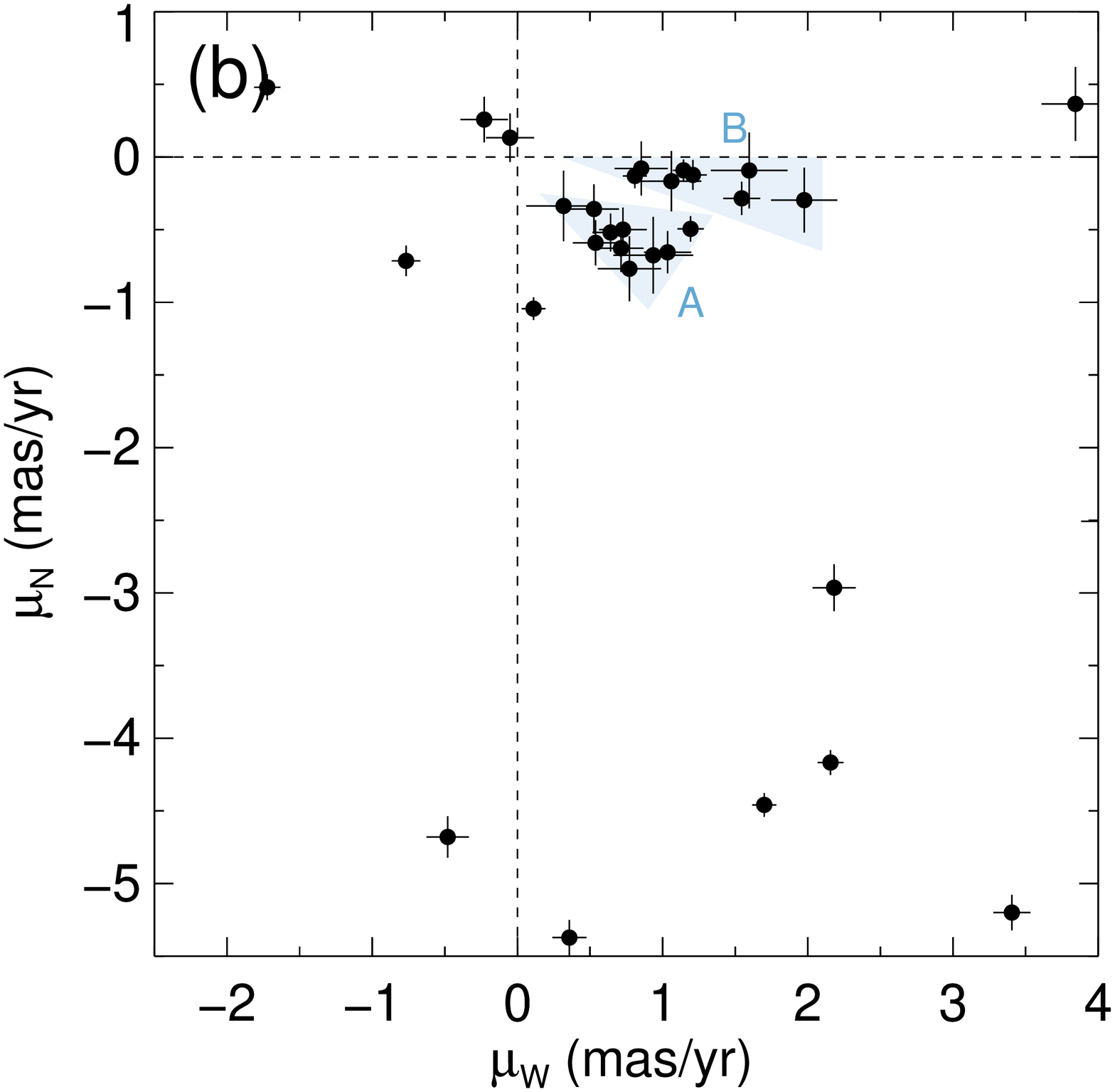}
\includegraphics[width=3.20in]{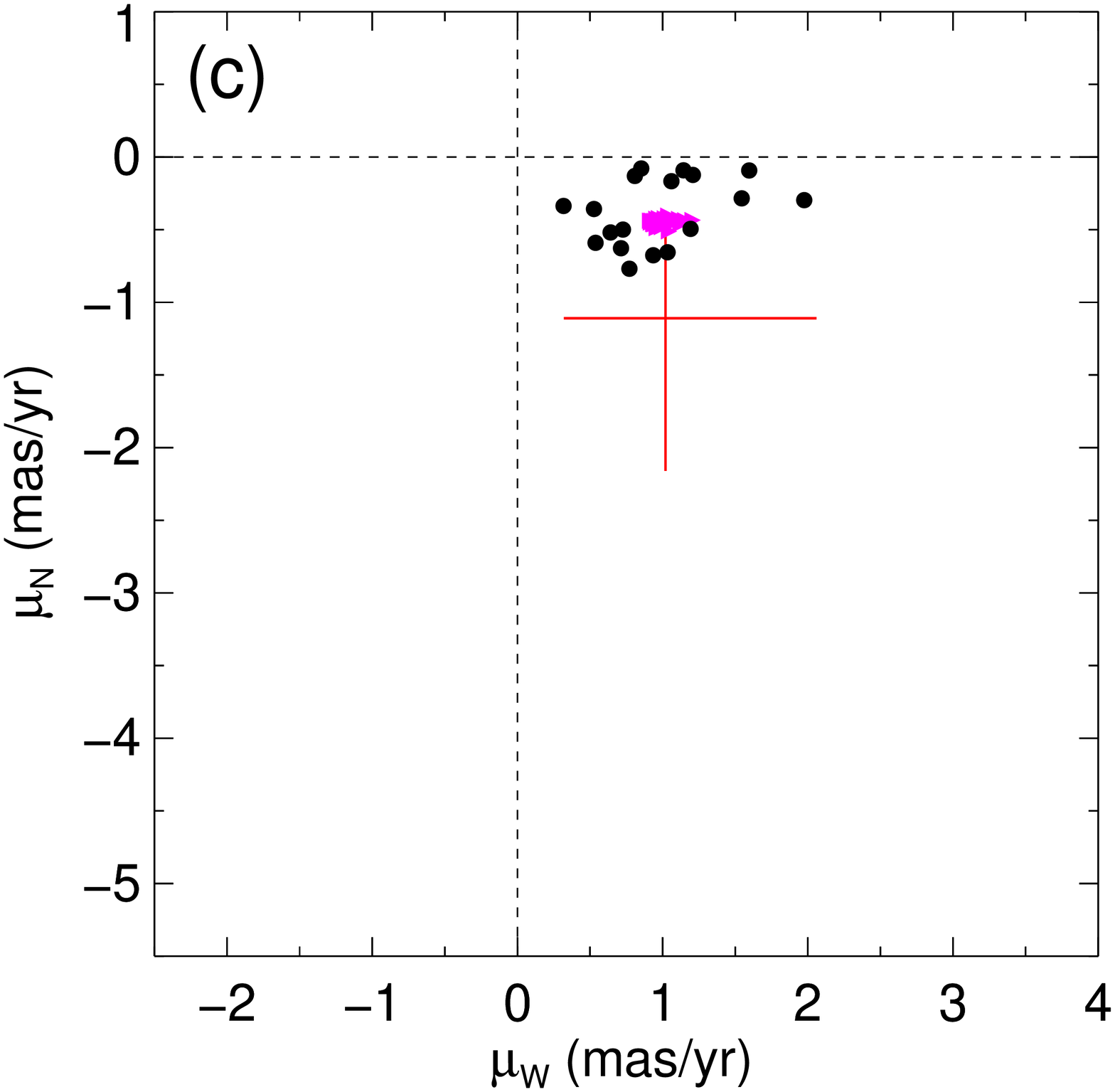}
\end{center}
\caption{Selection of stars associated with the Sgr stream in {\bf
  FIELD 3}. The panels and symbols are similar to those in
  Figure~\ref{fig:field1}. The group of magenta symbols in panel (c) 
  consists entirely of leading particles.
  An expectation value of 3 MW halo stars is predicted within the 
  area spanned by the red cross.
\label{fig:field3}}
\end{figure*}
%

%
\begin{deluxetable}{rcccc}
\tablecaption{Proper motions and photometry of Sgr stream 
stars in {\bf FIELD 3}\label{tab:field3}}
\tablehead{
 \colhead{}      & \colhead{$\muw$}     & \colhead{$\mun$}     & \colhead{F775W}     & \colhead{F606W}      \\
 \colhead{ID}    & \colhead{($\masyr$)} & \colhead{($\masyr$)} & \colhead{(VEGAMAG)} & \colhead{(VEGAMAG)} 
   }
\startdata
\cutinhead{Group A}
  1 & 1.03 $\pm$ 0.16 & $-$0.66 $\pm$ 0.15 & 21.04 $\pm$ 0.01 & 21.64 $\pm$ 0.01 \\
  2 & 0.71 $\pm$ 0.15 & $-$0.63 $\pm$ 0.16 & 23.04 $\pm$ 0.02 & 23.63 $\pm$ 0.03 \\
  3 & 1.19 $\pm$ 0.09 & $-$0.49 $\pm$ 0.09 & 23.18 $\pm$ 0.02 & 23.72 $\pm$ 0.12 \\
  4 & 0.64 $\pm$ 0.12 & $-$0.52 $\pm$ 0.13 & 23.96 $\pm$ 0.02 & 24.73 $\pm$ 0.02 \\
  5 & 0.53 $\pm$ 0.17 & $-$0.36 $\pm$ 0.17 & 24.07 $\pm$ 0.04 & 24.72 $\pm$ 0.03 \\
  6 & 0.73 $\pm$ 0.16 & $-$0.50 $\pm$ 0.15 & 24.31 $\pm$ 0.04 & 25.15 $\pm$ 0.03 \\
  7 & 0.54 $\pm$ 0.16 & $-$0.59 $\pm$ 0.16 & 24.32 $\pm$ 0.06 & 25.02 $\pm$ 0.04 \\
  8 & 0.32 $\pm$ 0.26 & $-$0.34 $\pm$ 0.24 & 24.81 $\pm$ 0.04 & 25.64 $\pm$ 0.01 \\
  9 & 0.94 $\pm$ 0.27 & $-$0.68 $\pm$ 0.26 & 25.03 $\pm$ 0.07 & 25.70 $\pm$ 0.08 \\
 10 & 0.77 $\pm$ 0.22 & $-$0.77 $\pm$ 0.22 & 25.07 $\pm$ 0.05 & 25.87 $\pm$ 0.13 \\
\cutinhead{Group B}
 11 & 1.14 $\pm$ 0.07 & $-$0.09 $\pm$ 0.07 & 21.18 $\pm$ 0.01 & 21.63 $\pm$ 0.08 \\
 12 & 1.21 $\pm$ 0.10 & $-$0.12 $\pm$ 0.10 & 21.48 $\pm$ 0.01 & 22.02 $\pm$ 0.02 \\
 13 & 1.54 $\pm$ 0.13 & $-$0.28 $\pm$ 0.11 & 21.89 $\pm$ 0.01 & 22.39 $\pm$ 0.01 \\
 14 & 0.81 $\pm$ 0.08 & $-$0.13 $\pm$ 0.08 & 23.39 $\pm$ 0.02 & 23.79 $\pm$ 0.21 \\
 15 & 1.06 $\pm$ 0.21 & $-$0.17 $\pm$ 0.21 & 24.12 $\pm$ 0.03 & 24.83 $\pm$ 0.03 \\
 16 & 1.60 $\pm$ 0.26 & $-$0.09 $\pm$ 0.26 & 24.67 $\pm$ 0.06 & 25.34 $\pm$ 0.03 \\
 17 & 1.98 $\pm$ 0.23 & $-$0.30 $\pm$ 0.22 & 24.73 $\pm$ 0.05 & 25.37 $\pm$ 0.07 \\
 18 & 0.85 $\pm$ 0.18 & $-$0.08 $\pm$ 0.19 & 25.18 $\pm$ 0.08 & 26.13 $\pm$ 0.08 
\enddata
\end{deluxetable}
%

Figure~\ref{fig:field3} shows the CMD, the PM diagram, and
the LM10 PM predictions for {\bf FIELD 3}. This field is located in
the leading arm of the Sgr stream, $\sim 73^{\circ}$ from the main
body of the Sgr dSph (see Figure~\ref{fig:fieldlocations}). So it is 
a sparse field, like {\bf FIELD 1}. However, it has a somewhat higher
density of foreground stars, as expected based on its Galactic
coordinates\footnote{$(l,b) = (341^{\circ},56^{\circ})$ for {\bf FIELD
    3}, versus $(166^{\circ},-61^{\circ})$ for {\bf FIELD 1}.}, and in
particular its Galactic longitude. Among all of the stars detected 
in this field, a total of 53 stars were considered for further 
analysis after rejecting stars with 1-D PM errors of $> 0.3\ \masyr$.
The adopted distance for the overlaid isochrones in the CMD is 
$57.8\ \kpc$; for comparison, the median distance of the LM10 model 
particles is $54.5\ \kpc$. The Sgr stream stars in {\bf FIELD 3} have 
the largest distances among those in all of our \hst\ fields, because 
this field is located near the apocenter of the leading arm.

We find that many stars in the CMD fall along one of our two fiducial
isochrones, and that a majority of these stars are located in the PM
diagram in a concentration that stretches roughly from $(\muw, \mun)
\approx (0.2, -1.0)\ \masyr$ to $(\muw, \mun) \approx (2.0,
0.0)\ \masyr$. The LM10 model predictions show a cold clump in more or
less the same part of PM space. These model particles are associated
with the leading-arm component, and were released from the main Sgr
body at 1.5--3 Gyr ago. The LM10 model does not predict trailing-arm
or secondary-wrap particles in {\bf FIELD 3} (because these particles
fall at different Sgr latitude $B_{\odot}$), and indeed, no obvious
additional cold clumps are detected in PM space. The stars
detected in other parts of the PM diagram are not clumped, and have
CMD properties that are inconsistent with the fiducial isochrones
(mostly stars with redder colors). So these are likely foreground
or background objects.

We identify 18 stars that cluster in PM space, and divide them
into two groups denoted A and B, as indicated in 
Figure~\ref{fig:field3}b. We list the PMs of individual stars in
these groups in Table~\ref{tab:field3}.  It is not obvious that A and
B would have to be physically distinct groups, but the PM diagram does
suggest the possible presence of two separate clumps. Colored circles
and labels in Figure~\ref{fig:field3}a indicate which star belongs to
which group.  We find no unique one-to-one association between the PM
groups and the OMP or IMR isochrones.

Group~B is well separated in PM space from the predicted 
location of MW halo stars. So we conclude from the combined CMD and
PM data that we have confidently detected the leading arm of the Sgr
stream. Most likely, group~A is also part of the leading arm of the
Sgr stream, but this is somewhat less certain. The mean PM of group~A
is close to the mean PM predicted for MW halo stars in this field.
Also, the number of stars in group~A is comparable to the predicted
number of MW halo stars that pass our CMD cuts (12 stars total, 6
stars in the area spanned by the red cross in 
Figure~\ref{fig:field3}c). However, group A is significantly colder
in PM space than what is predicted for any smooth MW halo
population. So either group~A is not a MW halo population, or the
actual properties of the MW halo must differ significantly
from what is assumed in the Besancon models.

\subsection{{\bf FIELD 4}}
\label{sec:field4}

%
\begin{figure*}
\begin{center}
\includegraphics[width=3.20in]{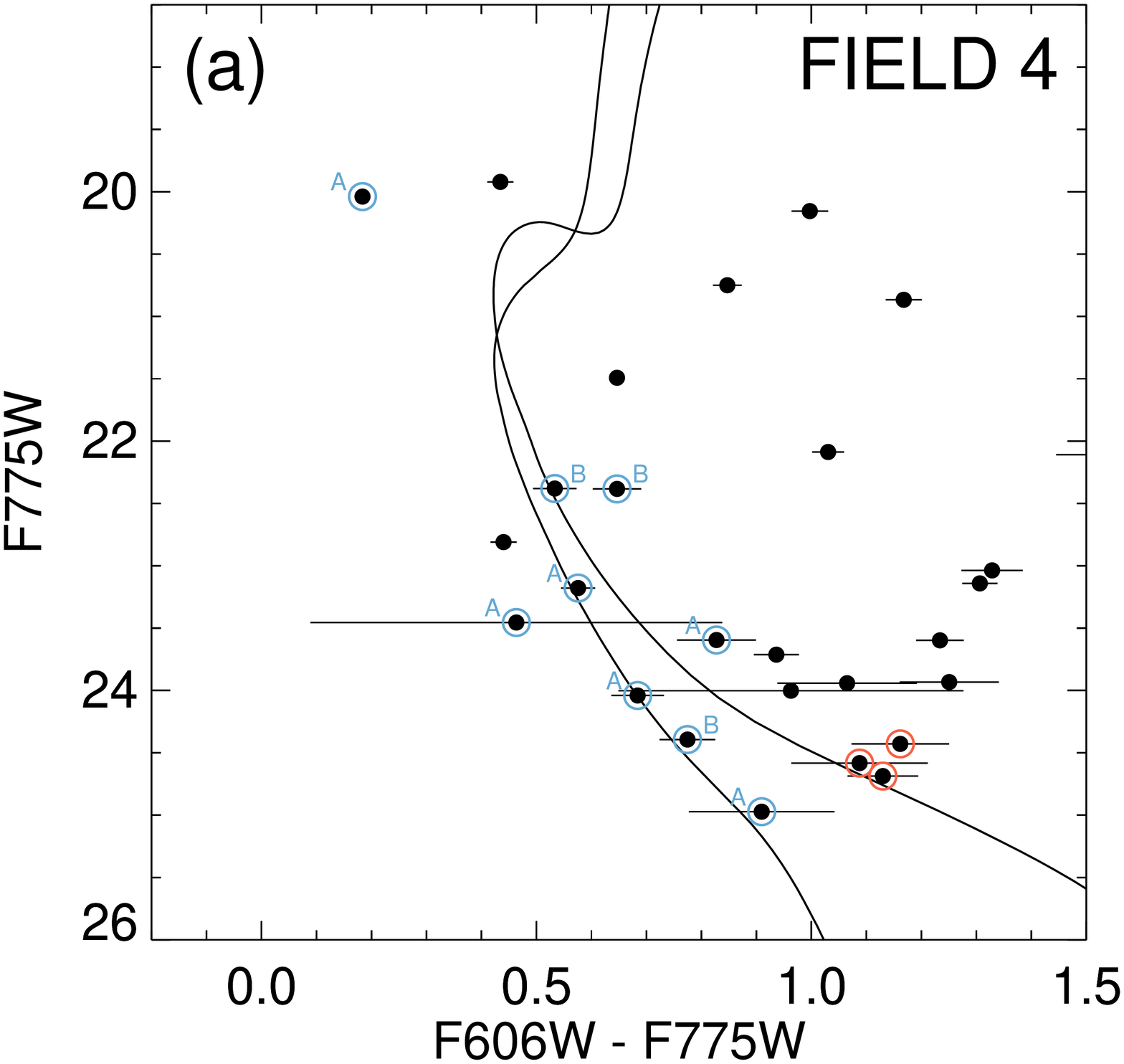}\\
\includegraphics[width=3.20in]{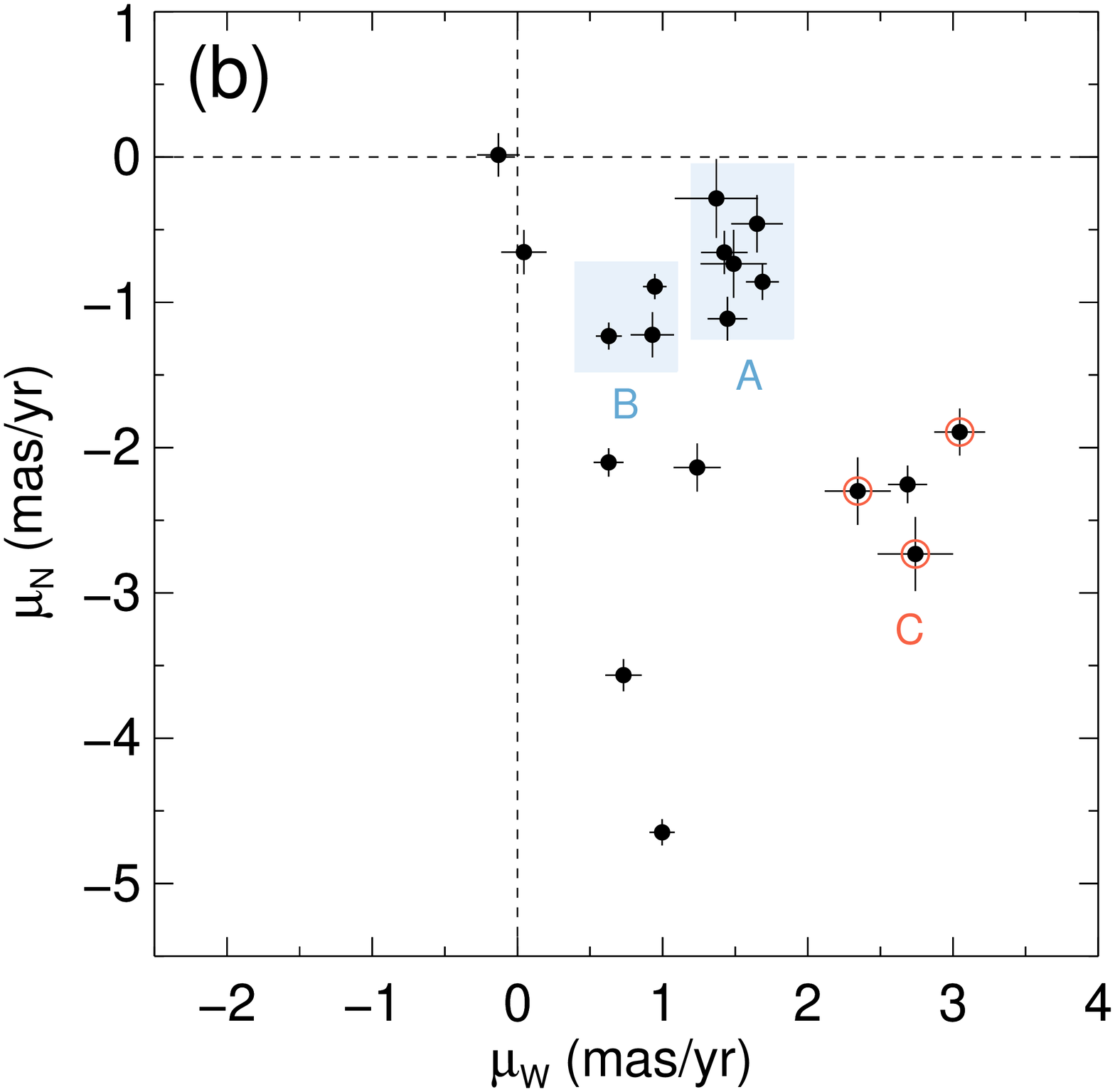}
\includegraphics[width=3.20in]{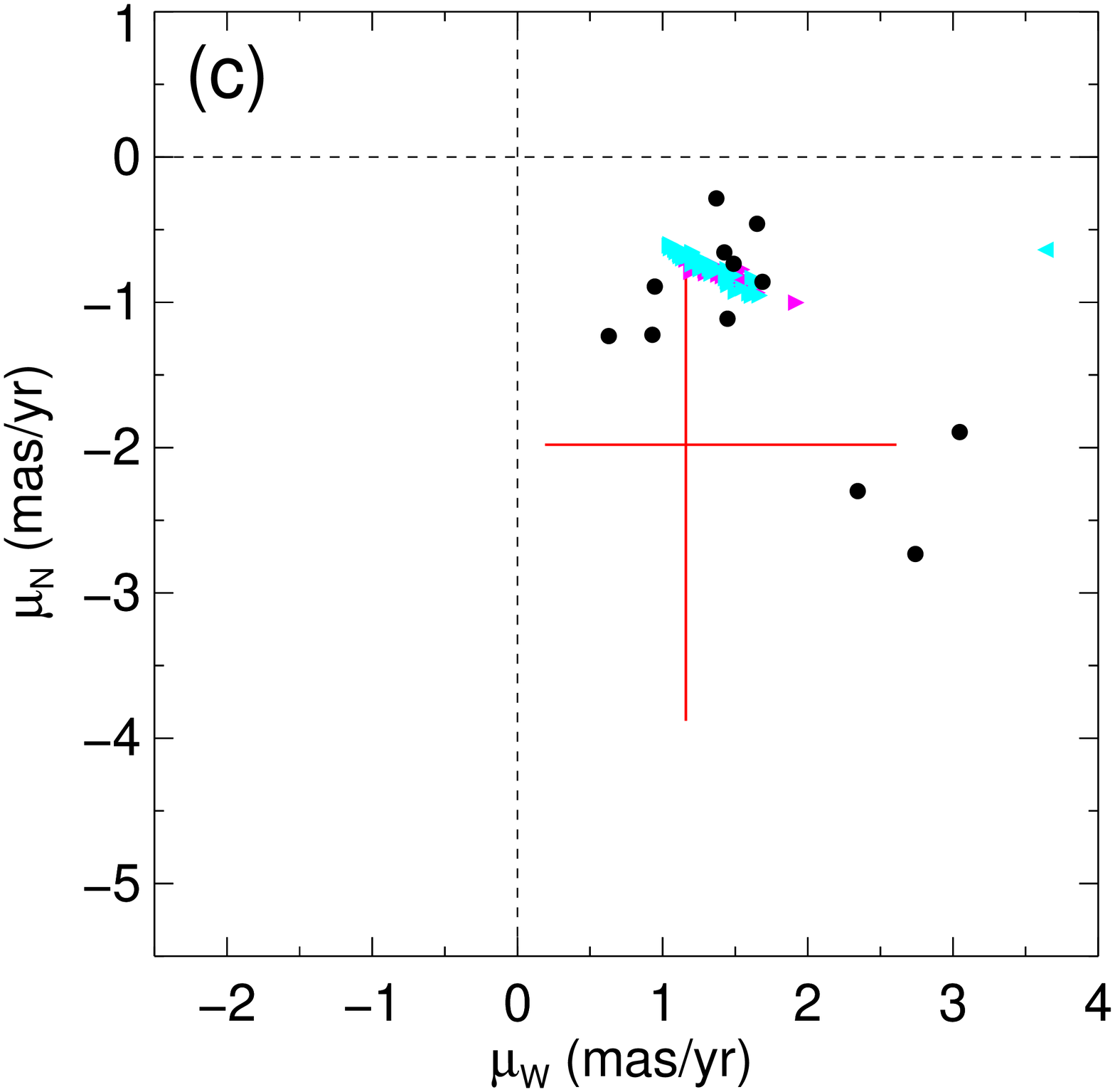}
\end{center}
\caption{Selection of stars associated with the Sgr stream in {\bf
 FIELD 4}. The panels and symbols are similar to those in
 Figure~\ref{fig:field1}. The group of cyan and magenta symbols
 in panel (c) consists entirely of leading particles, except
 for one trailing particle located far off to the right.
 An expectation value of 4 MW halo stars is predicted within the 
 area spanned by the red cross.
 \label{fig:field4}}
\end{figure*}


%
\begin{deluxetable}{rcccc}
\tablecaption{Proper motions and photometry of Sgr stream 
stars in {\bf FIELD 4}\label{tab:field4}}
\tablehead{
 \colhead{}      & \colhead{$\muw$}     & \colhead{$\mun$}     & \colhead{F775W}     & \colhead{F606W}      \\
 \colhead{ID}    & \colhead{($\masyr$)} & \colhead{($\masyr$)} & \colhead{(VEGAMAG)} & \colhead{(VEGAMAG)} 
   }
\startdata
\cutinhead{Group A}
  1 & 1.65 $\pm$ 0.18 & $-$0.46 $\pm$ 0.20 & 20.04 $\pm$ 0.01 & 20.22 $\pm$ 0.00 \\
  2 & 1.69 $\pm$ 0.11 & $-$0.86 $\pm$ 0.12 & 23.18 $\pm$ 0.03 & 23.75 $\pm$ 0.02 \\
  3 & 1.45 $\pm$ 0.14 & $-$1.11 $\pm$ 0.15 & 23.45 $\pm$ 0.04 & 23.92 $\pm$ 0.37 \\
  4 & 1.43 $\pm$ 0.16 & $-$0.66 $\pm$ 0.15 & 23.60 $\pm$ 0.03 & 24.42 $\pm$ 0.07 \\
  5 & 1.37 $\pm$ 0.29 & $-$0.28 $\pm$ 0.27 & 24.04 $\pm$ 0.05 & 24.72 $\pm$ 0.01 \\
  6 & 1.49 $\pm$ 0.23 & $-$0.73 $\pm$ 0.23 & 24.97 $\pm$ 0.06 & 25.88 $\pm$ 0.12 \\
\cutinhead{Group B}
  7 & 0.95 $\pm$ 0.08 & $-$0.89 $\pm$ 0.09 & 22.38 $\pm$ 0.01 & 22.91 $\pm$ 0.04 \\
  8 & 0.63 $\pm$ 0.09 & $-$1.23 $\pm$ 0.09 & 22.38 $\pm$ 0.01 & 23.03 $\pm$ 0.04 \\
  9 & 0.93 $\pm$ 0.15 & $-$1.22 $\pm$ 0.16 & 24.39 $\pm$ 0.04 & 25.17 $\pm$ 0.03 \\
\cutinhead{Group C}
 10 & 3.05 $\pm$ 0.17 & $-$1.89 $\pm$ 0.16 & 24.43 $\pm$ 0.04 & 25.59 $\pm$ 0.08 \\
 11 & 2.34 $\pm$ 0.23 & $-$2.30 $\pm$ 0.23 & 24.58 $\pm$ 0.06 & 25.67 $\pm$ 0.11 \\
 12 & 2.74 $\pm$ 0.26 & $-$2.73 $\pm$ 0.26 & 24.69 $\pm$ 0.05 & 25.82 $\pm$ 0.04
\enddata
\end{deluxetable}
%

Figure~\ref{fig:field4} shows the CMD, the PM diagram, and
the LM10 PM predictions for {\bf FIELD 4}. This field is located
further out in the leading arm of the Sgr stream, $\sim 109^{\circ}$
from the main body of the Sgr dSph (see
Figure~\ref{fig:fieldlocations}). So it is a sparse field, like {\bf
  FIELD 1} and {\bf FIELD 3}. Its density of foreground stars is
intermediate between those fields, as expected based on its Galactic
coordinates\footnote{$(l,b) = (258^{\circ},72^{\circ})$ for {\bf FIELD
 4}.}, and in particular its Galactic longitude. Among all of the 
stars detected in this field, a total of 30 stars were considered for 
further analysis after rejecting stars with 1-D PM errors of 
$> 0.3\ \masyr$. The adopted distance for the overlaid isochrones in 
the CMD is $37.8\ \kpc$; for comparison, the median distance of the 
LM10 model particles is $38.3\ \kpc$.

Inspection of the CMD shows that many stars fall along one of our two
fiducial isochrones. Most of the stars consistent with our fiducial
isochrones are located in the PM diagram in a concentration around
$(\muw, \mun) \approx (1.3, -0.8)\ \masyr$. The LM10 model particles
cluster in more or less the same part of PM space. These model
particles are associated with the leading-arm component, and most of
them were released from the main Sgr body at 3--5 Gyr ago. This is
1--2 Gyr earlier than the stars in {\bf FIELD 3}, which is why these
stars are now found at a larger angular distance from the Sgr dSph.

We focus on 8 stars observed near the leading arm PM predicted
by LM10. Based on the apparent clustering in the PM diagram, we
divide these stars into two groups. We denote these groups as A and
B, as indicated in Figure~\ref{fig:field4}b. The PMs of the individual
stars in these groups are listed in Table~\ref{tab:field4}. As are
the cases of {\bf FIELDS 1} and {\bf 3}, it is not obvious that A and
B would have to be physically distinct groups, but the PM diagram does
suggest the possible presence of two separate clumps. Colored circles
and labels in Figure~\ref{fig:field4}a indicate which star belongs to
which group. There is no unique one-to-one association between the PM
groups and the OMP or IMR isochrones.

Most of the stars detected in other parts of the PM diagram are
not clumped, and have CMD properties that are inconsistent with the
fiducial isochrones (mostly stars with redder colors). So these are
likely foreground or background objects. However, there is some
evidence for an additional clump of stars in PM space, which we have
labeled ``C'' in Figure~\ref{fig:field4}b. These stars could
represent a coherent cold component, although this cannot be
established with statistical significance given the low number of
stars.

Group~A is well separated in PM space from the predicted
location of MW halo stars. So we conclude from the combined CMD and
PM data that we have confidently detected the leading arm of the Sgr
stream. Whether group~B is also part of the Sgr stream is not clear.
The mean PM of group~B is close to the mean PM predicted for MW halo
stars in this field. Also, the number of stars in group~B is
comparable to the predicted number of MW halo stars that pass our
CMD cuts (4 stars in the area spanned by the red cross in
Figure~\ref{fig:field4}c). The low number of stars in group~B, and
the presence of additional stars in the PM diagram which may or may
not be associated, makes it difficult to determine whether the PM
dispersion of this group is or is not consistent with that predicted
for a MW halo population.

Group~C could indicate the presence of an unrelated stream in
the MW halo, given that the PM of this clump is not consistent with
any component seen in the LM10 model. Trailing stream particles at
this Sgr longitude $\Lambda_{\odot}$ in the LM10 model are mostly
located at different latitude $B_{\odot}$ than {\bf FIELD 4} (although
one such particle is seen in Figure~\ref{fig:field4}c) and have PMs
that cluster near $(\mu_W,\mu_N) \approx (4.0,-0.5)$. The median
distance of these trailing-arm stars is $21.4\ \kpc$, which is closer
than the distance of the leading-arm particles. The CMD properties of
the C clump stars (circled red in Figure~\ref{fig:field4}a) are not
inconsistent with this. So we cannot rule out that the C clump does in
fact represent the trailing arm of the Sgr stream. If so, then
the LM10 model does not correctly represent the dynamics of these
particles.  There is in fact some independent evidence that this may
be the case.  Specifically, \cite{bel14} showed that in the trailing
arm of the stream, at angular distances $\gta 150^{\circ}$ from the
Sgr dSph, the LM10 model may not correctly reproduce the observed
stream distances. In view of this, we cannot at present uniquely
identify the nature of the C clump stars in {\bf FIELD 4}.

%
\begin{deluxetable}{lccccc}
\tablecaption{Proper Motion Average and Dispersion for Sgr Stars\label{tab:pmresults}}
\tablehead{
 \colhead{}       & \colhead{$\muw$}     & \colhead{$\mun$}     & \colhead{}                             & \multicolumn{2}{c}{$\sigma$\tablenotemark{b}} \\
 \cline{5-6}
 \colhead{Sample} & \colhead{($\masyr$)} & \colhead{($\masyr$)} & \colhead{$N_{\star}$\tablenotemark{a}} & \colhead{($\masyr$)} & \colhead{($\kms$)}
}
\startdata
\sidehead{{\bf FIELD 1} ($N_{\rm gal}$\tablenotemark{d}$ = 246$)}
Group A   & 0.20 $\pm$ 0.11 & $-$1.68 $\pm$ 0.04 & \phn5 & 0.19 &    \phn30.1 \\
Group B   & 0.10 $\pm$ 0.07 & $-$2.47 $\pm$ 0.08 & \phn5 & 0.05 & \phn\phn8.5 \\ 
Group C   & 0.69 $\pm$ 0.18 & $-$4.25 $\pm$ 0.30 & \phn3 & 0.40 &    \phn62.7 \\
A$+$B     & 0.14 $\pm$ 0.07 & $-$2.06 $\pm$ 0.12 &    10 & 0.28 &    \phn44.8 \\
A$+$B$+$C & 0.27 $\pm$ 0.09 & $-$2.50 $\pm$ 0.25 &    13 & 0.68 &       107.2 \\
\hline
\sidehead{{\bf FIELD 2} ($N_{\rm gal}$\tablenotemark{d}$ = 25$)}
All       & 2.89 $\pm$ 0.07 & $-$1.40 $\pm$ 0.11 &    90 & 0.19 &    \phn27.3 \\
\hline
\sidehead{{\bf FIELD 3} ($N_{\rm gal}$\tablenotemark{d}$ = 141$)}
Group A   & 0.83 $\pm$ 0.12 & $-$0.54 $\pm$ 0.04 &    10 & 0.12 &    \phn32.5 \\
Group B   & 1.13 $\pm$ 0.12 & $-$0.14 $\pm$ 0.04 & \phn8 & 0.21 &    \phn58.2 \\
A$+$B     & 0.98 $\pm$ 0.10 & $-$0.38 $\pm$ 0.05 &    18 & 0.29 &    \phn79.9 \\  
\hline
\sidehead{{\bf FIELD 4} ($N_{\rm gal}$\tablenotemark{d}$ = 169$)}
Group A   & 1.55 $\pm$ 0.06 & $-$0.77 $\pm$ 0.10 & \phn6 & 0.12 &    \phn21.9 \\ 
Group B   & 0.82 $\pm$ 0.10 & $-$1.07 $\pm$ 0.11 & \phn3 & 0.14 &    \phn24.8 \\ 
Group C\tablenotemark{c}   & 
            2.78 $\pm$ 0.18 & $-$2.18 $\pm$ 0.21 & \phn3 & 0.26 &    \phn46.0 \\ 
A$+$B     & 1.29 $\pm$ 0.11 & $-$0.83 $\pm$ 0.10 & \phn9 & 0.30 &    \phn54.1
\enddata
\tablenotetext{a}{Number of Sgr stream stars included in the PM calculations.}
\tablenotetext{b}{Average one-dimensional velocity dispersion, obtained from 
the combined West and North measurements, corrected for observational
scatter by subtracting the median random PM error bar in quadrature. The 
transformations from $\masyr$ to $\kms$ are based on the same distances as 
used for the isochrones in Figures~\ref{fig:field1}a--\ref{fig:field4}a.}
\tablenotetext{c}{The group C stars in {\bf FIELD 4} are not confirmed as part
of the Sgr stream, but they could be.}
\tablenotetext{d}{Number of background galaxies used to set the zero point of 
our PMs for each field.}
\end{deluxetable}
%

\section{Proper Motion Variation Along the Stream}
\label{sec:pmvar}

\subsection{Average HST Proper Motions}
\label{sec:avgmotions}

The analysis of Sections~\ref{sec:data} and~\ref{sec:sgrstream} has
yielded a list of Sgr stream stars in each \hst\ field 
(Tables~\ref{tab:field1}--\ref{tab:field4}). Using these lists, we 
derived the average PM of the Sgr stars in each field, and its 
associated uncertainty. We did this separately for each identified 
subgroup in each field (A, B, or C). We also did this for the 
combined A+B subgroups in {\bf FIELDS 1, 3, 4}, since it is not 
obvious for these fields whether the distinction between groups A and 
B is in fact statistically significant or physically meaningful. 
For {\bf FIELD 1} we also list the average for
the combined A+B+C sample, which includes all identified Sgr stream
stars in that field.

The results are presented in Table~\ref{tab:pmresults}. When
calculating average PMs for an individual sample (A, B, or C), we
adopted the error-weighted mean. The error of this mean was computed
through the bootstrapping method \citep{efr93}. This implies that the
uncertainties are ultimately based on the scatter between the
data points, and not merely on propagation of the formal per-star
random PM uncertainties. So this takes into account any intrinsic PM
dispersion between stars, as well as any possible remaining
systematic in our measurements. For the combined samples (A+B or
A+B+C) we adopted the un-weighted mean (and the standard
error-in-the-mean to define the uncertainty), given that the scatter
for these samples is not dominated by measurement uncertainties, but
by kinematical differences between populations.

We also list for each sample in Table~\ref{tab:pmresults} an estimate
of the average intrinsic one-dimensional dispersion $\sigma$
transverse to the line of sight, in both $\masyr$ and $\kms$. The
listed values combine the measurements in the West and North
directions, with measurement errors subtracted in quadrature. 
For the individual samples (A, B, or C), we find $\sigma$ values 
in the range $9-63\ \kms$. Our measured dispersions are on average  
higher than the known range of LOS velocity dispersions 
$\sigma_{\rm LOS} = 8-27\ \kms$ \citep{maj04,mon07,car12,kop13}.
This probably reflects, at least in part, true intrinsic scatter 
and not merely unquantified measurement errors. The velocity 
dispersion of the stream is largest in the direction along the 
stream, which is primarily sampled by the PM measurements. 
By comparison, LOS measurements primarily sample the velocity 
dispersion perpendicular to the stream, which tends to be smaller. 

For the combined samples (A+B or A+B+C) the measured velocity
dispersions tend to be larger than for the individual samples,
reflecting primarily the kinematical differences between the samples.

We also verified that the numbers of stars ($N_*$) identified 
kinematically in Table~\ref{tab:pmresults} as potential stream stars 
are reasonable, given our existing understanding of the Sgr stream 
from photometric studies. To this end, we downloaded lists of sources 
classified as stars in the SDSS database within a $1\times1$ deg$^2$
region around our target field locations. All SDSS photometry was then
de-reddened for comparison with theoretical isochrones. We used
isochrones with the same age, [Fe/H], [$\alpha$/H], and distance as the
fiducial isochrones we used in our study (i.e., old metal-poor and
intermediate-age metal-rich). We then counted the number of stars
consistent with the corresponding populations down to the 95\%\ 
limiting magnitude of the SDSS photometry. To allow subtraction of 
fore- and backgrounds for {\bf FIELD 1}, we downloaded a list of stars 
in a $1\times1$ deg$^2$ region located at the same Galactic longitude 
but at the opposite Galactic hemisphere. We then counted the number of 
stars using the same isochrones, and subtracted this number from the
original counts. For {\bf FIELDS 3} and {\bf 4}, SDSS does not have 
coverage in the opposite Galactic hemisphere, so for those we picked 
instead the average of fields above and below our target fields in 
Galactic coordinates. After completion of the fore- and background 
subtraction, we used theoretical Salpeter-type luminosity functions to 
estimate the number of stars in the magnitude range of our \hst\ 
observations. Finally, this number was scaled down to the predicted
number of stars in the area of the ACS/WFC field of view, by 
multiplying by a factor of 
$(202\arcsec\times202\arcsec)/(3,600\arcsec\times3,600\arcsec)$.

Using the method above for {\bf FIELD 1}, we get estimates of 8 stars 
for the older population and 4 stars for the younger population, i.e., 
12 stars total. This is very close to the number of 13 Sgr stream stars
identified kinematically in Table~\ref{tab:field1}. For {\bf FIELDS 3} 
and {\bf 4}, we get 13 (older) + 6 (younger) = 19 stars and
3 (older) + 2 (younger) = 5 stars, respectively. These can be
compared to the total numbers of stars reported in Groups~A and B of
Tables~\ref{tab:field3} and \ref{tab:field4}, namely 18 and 9, 
respectively. While the agreement is not perfect, it is adequate given 
the small-number statistics and the uncertainties in the estimation 
process. In summary, for all \hst\ fields, the number of kinematically 
identified Sgr stream stars is consistent with the extrapolation of 
photometric results from SDSS. This strengthens our conclusion that 
the kinematically identified populations are indeed part of the 
Sagittarius stream.

\subsection{Comparison to Other Proper Motion Measurements}
\label{sec:PMcomparison}

\subsubsection{Trailing Arm}
\label{sec:trailPM}

\begin{figure}
\epsscale{1.1}
\plotone{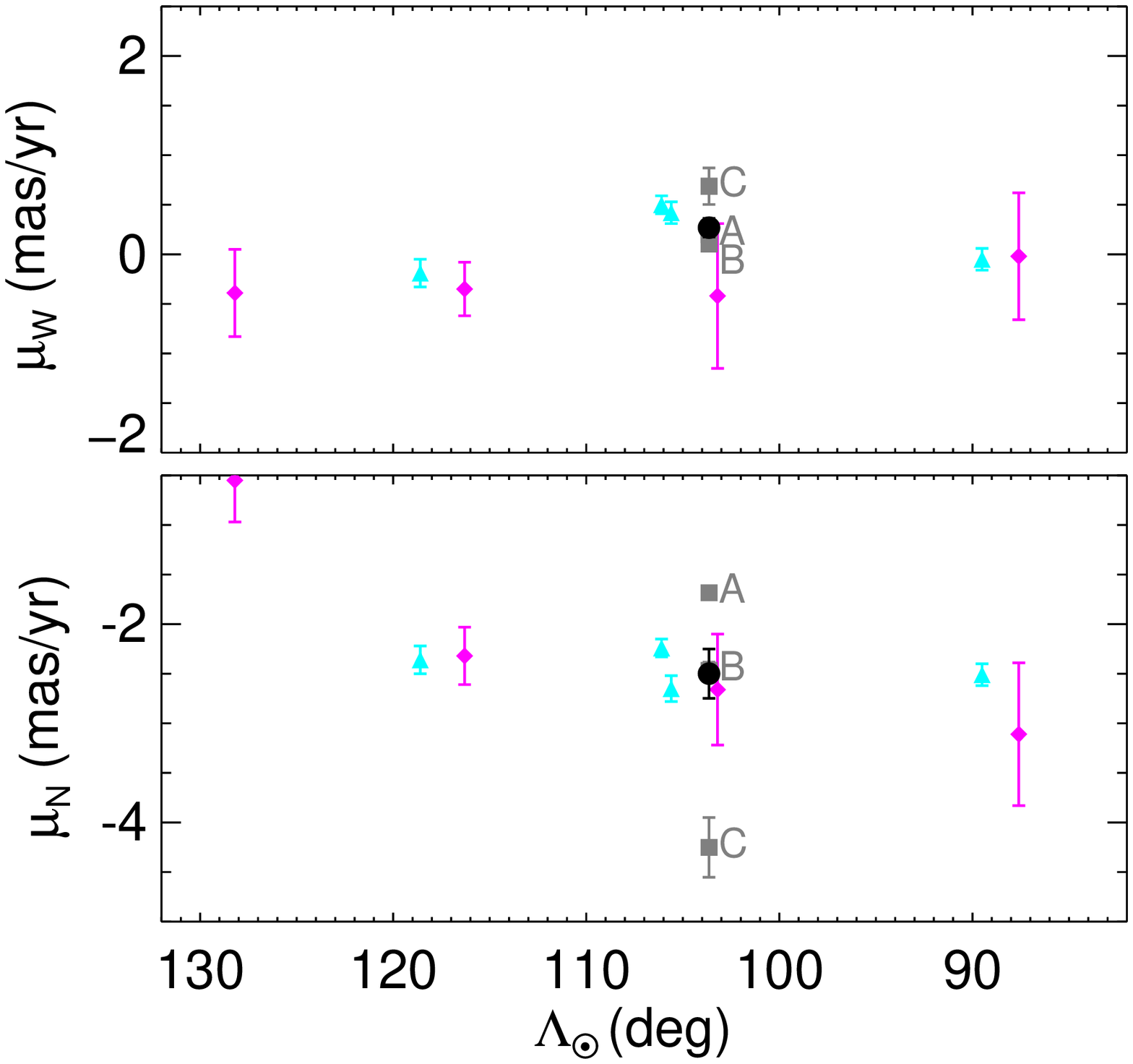}
\caption{Proper motions in the trailing arm, in the west (upper
  panel) and north (lower panel) directions, as a function of the Sgr
  latitude. Ground-based measurements from \citet[][magenta 
  diamonds]{car12} and \citet[][cyan triangles]{kop13} are shown for 
  comparison. Our \hst\ average for all identified Sgr stream stars in 
  {\bf FIELD 1} (the combined A+B+C sample, as defined in 
  Figure~\ref{fig:field1} and Table~\ref{tab:pmresults}) is shown in 
  black circle. Overall, the studies are in good agreement. The average 
  PMs of our {\bf FIELD 1} A, B, and C samples are shown in grey squares.
\label{fig:comp_field1}}
\end{figure}

The only PMs previously reported for the Sgr stream were obtained 
from ground-based observations of the trailing part of the stream.
\citet{car12} identified stream stars in 4 fields located between
$\Lambda_{\odot} = 70$--130 \degr. They obtained average PMs of 
spectroscopically confirmed Sgr stream member stars with random 
errors per coordinate ranging from 0.3--1.0 $\masyr$, from 
observations collected with photographic plates over an 
$\sim 80$ year time baseline.
\citet{kop13} instead identified stream stars in 3 fields in
the same Sgr longitude range, from observations of the Stripe 82
region from the SDSS. They obtained average PMs of thousands of 
Sgr stream stars with lower random errors of $\sim 0.1\ \masyr$, 
from observations collected over a $\sim 7$ year time baseline. 
In Figure~\ref{fig:comp_field1}, we show the measured PMs along 
the west (upper panel) and north (lower panel) directions, as a 
function of the Sgr longitude $\Lambda_{\odot}$.

The \hst\ measurements for our trailing arm {\bf FIELD 1} are shown
for comparison in Figure~\ref{fig:comp_field1}. \hst\ {\bf FIELD 1} 
is located at a similar $\Lambda_{\odot}$ as the SA93 field of 
\citet{car12}, although there is a $\sim 7\degr$ difference in Sgr 
latitude $B_{\odot}$ between the fields. \hst\ {\bf FIELD 1} is also 
located at a similar $\Lambda_{\odot}$ as the FP2 and FS4 
measurements of \citet{kop13}, and with only a $\sim 3\degr$ 
difference in $B_{\odot}$. The FP2 and FS4 measurements correspond 
to similar field locations, but with the Sgr stars selected using 
different criteria (without or with spectroscopic information, 
respectively).

When comparing to the ground-based measurements, it is most
appropriate to use the A+B+C sample average (as defined in
Figure~\ref{fig:field1} and Table~\ref{tab:pmresults}). This includes
{\it all} Sgr stream stars, independent of their PMs. This yields a
large PM uncertainty, because we are averaging stars that do not 
belong to the same kinematical population. This is appropriate for 
comparison to ground-based projects, since those could not separate 
different kinematical populations based on their PMs. However, in any 
dynamical modeling it would obviously be best to model the different 
kinematical populations, which individually have much smaller errors 
bars, separately. We find our A+B+C results to be consistent with 
those for the \citet{car12} field SA93 at the $\sim 1\sigma$ level. 
Our results are also consistent with those for the \citet{kop13} FP2 
and FS4 samples at roughly the $1\sigma$ level in $\muw$. In $\mun$, 
our A+B+C sample average falls between the FP2 and FS4 measurements, 
which themselves differ at the $\sim 2.5\sigma$ level. So overall, our
measurements are consistent with the available ground-based 
measurements. Also they fall roughly along the general trend with Sgr
longitude defined by the \citet{car12} and \citet{kop13} measurements.

These comparisons suggest that none of the PM studies suffers from
large, unquantified systematics. The ground-based data provide better
sampling of the longitude dependence of the PMs within the trailing
part of the stream. By contrast, the \hst\ data provide only a single
trailing-arm field. However, within this field we get smaller overall
uncertainties than the ground-based measurements, and a higher
information content. Specifically, for the ground-based data only the
average PM of all Sgr stream stars in each sample was measured. By
contrast, for our \hst\ data, we actually have accurate PM
measurements for {\it individual} stream stars. This allows us to
kinematically separate the different (trailing/leading) arms stars
based on their PMs, as shown in Figure~\ref{fig:field1}.

%
%

\subsubsection{Sgr dSph}
\label{sec:dSphPM}

The \hst\ {\bf FIELD 2} is located at $(\Lambda_{\odot},B) =
(-3.53^{\circ},4.65^{\circ})$. This places the Sgr stars in this field
at $5.8^{\circ}$ from the Sgr dSph. In $N$-body simulations, bound
material is found out to $\sim 8^{\circ}$ from the Sgr dSph 
\citep[e.g.,][]{law05}. Therefore, the Sgr stars identified in 
{\bf FIELD 2} give information primarily about the PM of the Sgr dSph 
itself, and less so about the kinematics of unbound stream material.

Several measurements of the PM of the Sgr dSph already exist in the
literature, based on different kinds of data: \citet{din05} used the
ground-based Southern Proper Motion Catalog 3; \citet{pry10} analyzed
\hst\ data of three fields; and \citet{mas13} analyzed \hst\ data of a
foreground globular cluster (NGC 6681), qualitatively similar to the
case for our {\bf FIELD 2}. Even though we are not the first to have
used \hst\ data to measure the PM of the Sgr dSph, our work does have
several advantages over the previous studies. \citet{pry10} used
foreground Galactic stellar populations to set the astrometric
reference frame, which requires assumptions about the PM kinematics of
the foreground Galactic stellar population to obtain absolute PMs.
\citet{mas13} instead used stationary background galaxies to set the
astrometric reference frame as we do here, but they used only 5
background galaxies, which were fitted as point sources. By contrast,
we used 24 background galaxies for our {\bf FIELD 2}, and for each of
these we build an individual template that takes the exact galaxy
morphology into account.

All of the individual PM estimates, including the one presented here,
pertain to different fields that do not coincide with the
center-of-mass (COM) of the Sgr dSph. This implies that two effects
need to be taken into account in any comparison. The first effect is
that possible internal motions, such as rotation, could in principle
contribute to the measurements for the different fields. However, this
should not be a problem for the Sgr dSph, since this galaxy does not
show any evidence of rotation \citep{pen11,fri12}. The second effect 
is that even in the absence of internal motions, one would not expect 
to measure the same PM for different fields, because of perspective 
effects. Depending on where one points in the Sgr dSph, different 
components of the 3D COM velocity vector project onto the local LOS, 
West, and North directions \citep{vdm02}.

After correcting for viewing perspective following \citet{vdm08} and 
\citet{mas13}, the following COM PM estimates are obtained for 
$(\muw,\mun)$: $(2.83 \pm 0.20, -1.56 \pm 0.20)\ \masyr$ for 
\citet{din05}; $(2.37 \pm 0.20, -1.63 \pm 0.22)\ \masyr$ for 
\citet{pry10}; $(2.56 \pm 0.18, -1.29 \pm 0.16)\ \masyr$ for 
\citet{mas13}. These measurements are mutually consistent, given the 
uncertainties, with a weighted average of $(2.59 \pm 0.11, -1.49 \pm 
0.11)\ \masyr$. The perspective-corrected estimate based on our 
\hst\ {\bf FIELD 2} data is $(2.92 \pm 0.07, -1.56 \pm 0.11)\ \masyr$.

The \hst\ {\bf FIELD 2} result is very close to the value implied by
the \citet{din05} measurement, but the agreement with the
\citet{pry10} ($2.6\sigma$ different in the West direction) and the
\citet{mas13} result ($1.9\sigma$ and $1.4\sigma$ different in the
West and North directions, respectively) is less good. When comparing
the \hst\ {\bf FIELD 2} result to the weighted average of the
previously published results, the difference is $(0.33 \pm 0.13, -0.07
\pm 0.16)\ \masyr$. Such a residual can occur by chance in a
two-dimensional Gaussian distribution at 4\% probability. Hence, it is
more likely that one or more of the measurements contain unquantified
systematics.  One way to account for this is to multiply all the error
bars by a factor $1.8$, which makes the results statistically
consistent ($\chi^2$ equal to the number of degrees of freedom). If we
do this, and then take the weighted average of all four measurements,
we obtain $(\muw,\mun) = (2.82 \pm 0.11 , -1.51 \pm 0.14)\ \masyr$.
This is our current best estimate estimate of the COM PM of the Sgr
dSph, based on all available measurements.

\subsection{Preliminary Comparison to Model Proper Motions}
\label{sec:PMmodel}

%
\begin{figure*}
\epsscale{1.15}
\plotone{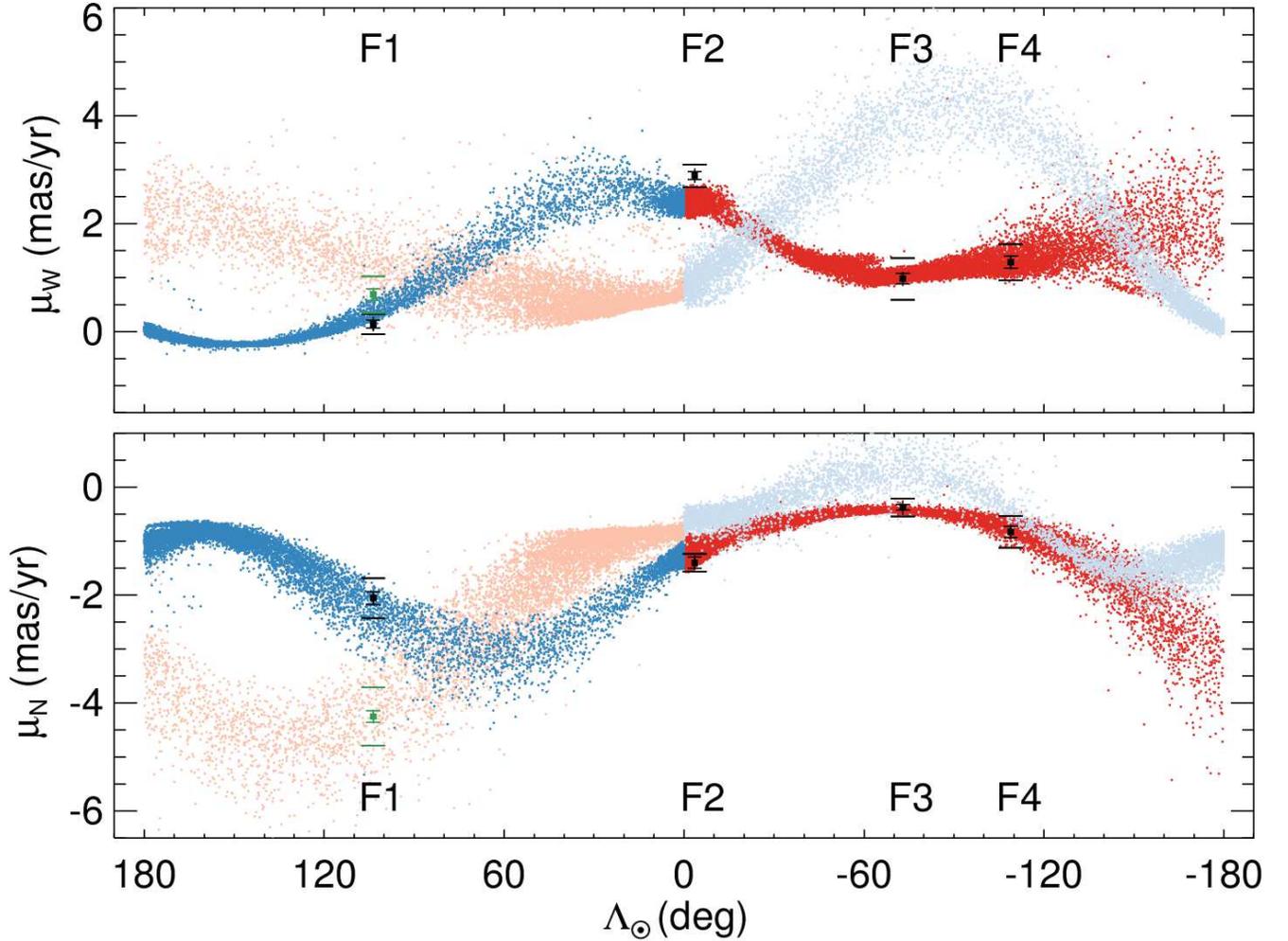}

\caption{Comparison of \hst\ PM observations of $\muw$ (top) and
  $\mun$ (bottom) to the PMs of the $N$-body particles in the LM10
  model, as function of the Sgr longitude $\Lambda_{\odot}$. The 
  horizontal axis is chosen so that the Sgr dSph lies in the middle of 
  each panel. The leading arm (red) extends to right from the dSph, 
  while   the trailing arm (blue) extends to the left (as is true for 
  the view in equatorial coordinates; compare
  Figure~\ref{fig:fieldlocations}). The primary wrap at each longitude
  (i.e., particles within $180^{\circ}$ in Sgr longitude from the
  dSph) are shown in bold colors. The secondary wrap at each longitude
  (i.e., particles that are between $180^{\circ}$--$360^{\circ}$ in
  Sgr longitude from the dSph) are shown with lighter
  colors. Particles that have an orbital phase that differs by more
  than $360^{\circ}$ from that of the dSph are not shown. A black
  solid dot with error bars shows for each \hst\ field the average
  observed PM (from Table~\ref{tab:pmresults}) for the stars
  identified as belonging to the primary wrap. This uses the A+B
  subsample averages for {\bf FIELD 1, 3, and 4}. For {\bf FIELD 1} we
  also show as a green square the subsample C, identified as
  corresponding to the secondary (leading) wrap at that Sgr
  longitude. Horizontal bars above and below the data points indicate
  $\pm$ the PM dispersions estimated from the measurements (calculated
  as in Table~\ref{tab:pmresults}, but now for the West and North
  directions separately).
\label{fig:pmresults}}
\end{figure*}
%

For a preliminary assessment of what the new \hst\ PM data may imply
for our understanding of the Sgr stream, we qualitatively compare the
data to the PM predictions of the existing $N$-body models of 
LM10 and \citet{pen10}. 
Figure~\ref{fig:pmresults} shows the PMs $\muw$ and $\mun$ of the
LM10 model's $N$-body particles (colored) as function of the Sgr 
longitude $\Lambda_{\odot}$. The \hst\ PM averages (black) are 
overplotted for comparison. Overall, the new measurements follow quite 
closely the predicted PM trend with Sgr longitude. This constitutes a 
remarkable success for the LM10 model. The model was fit only to 
distance and LOS velocity data for the stream, with no reference to 
PMs. So Figure~\ref{fig:pmresults} does not represent a fit to the 
data, but rather a successful verification of a prediction that was 
made {\it a priori}. Without further quantitative comparison, our PM 
results do not necessarily imply that the LM10 model is the only 
correct interpretation of the data, or that indeed the MW halo must be
triaxial in the manner suggested by LM10. However, our new \hst\ PM
data certainly provide no immediate evidence for an obvious problem
with the model.

Figure~\ref{fig:pmresults} shows that the biggest mismatch between the
LM10 model and the \hst\ PM data occurs for {\bf FIELD 2}, which is
sensitive primarily to the PM of the Sgr dSph. In the LM10 model, the
best-fit Galactocentric velocity for the Sgr dSph was found to be
$(V_X , V_Y , V_Z) = (230, -35, 195)\ \kms$. With the same assumptions
made by LM10 for the solar circular velocity ($V_{\rm circ} = 
  220\ \kms$) and Sgr dSph distance ($D = 28\ \kpc$), the COM PM
predicted by the model is $(\muw,\mun) = (2.45, -1.30)\ \masyr$. This
is in the same direction on the sky as the best-estimate PM derived in
Section~\ref{sec:dSphPM}, but lower in amplitude by about 14\%. This
discrepancy may be resolved by one or a combination of several
things. 

\citet{car12} constructed variations to the $N$-body models of LM10, 
in which $V_{\rm circ}$ was treated as a free parameter. They showed 
that when $V_{\rm circ}$ is increased from the canonical $220\ \kms$ 
to a value as high as $264\ \kms$, then the best-fit PM of the Sgr 
dSph (see their Figure~19) is $(\muw,\mun) = (2.78, -1.55)\ \masyr$. 
This is in excellent agreement with our best estimate estimate COM PM
$(\muw,\mun) = (2.82 \pm 0.11 , -1.51 \pm 0.14)\ \masyr$, based on 
all available PM measurements. However, it is not obvious that such
a large $V_{\rm circ}$ is plausible in the context of other 
astronomical knowledge. The azimuthal velocity of the sun in this
model, taking into account also the $12\ \kms$ solar peculiar 
azimuthal velocity, is $v_{\phi,\odot} = 276\ \kms$. \citet{bov12} 
recently found from a detailed study of APOGEE data that 
$v_{\phi,\odot} = 242^{+10}_{-3}\ \kms$. Also, \cite{car12} found 
from fitting their trailing arm PM data a best fit value 
$V_{\rm circ} = 232 \pm 14\ \kms$. So while there is now growing 
consensus that the solar velocity may be larger than previously 
believed (and adopted by LM10), its value may not be quite as large 
as needed to bring the $N$-body models in agreement with the Sgr 
dSph PM measurements.

Another change to the LM10 model that would bring its PM
predictions closer to the measurements, would be to decrease the 
distance of the Sgr dSph. For example, excellent agreement would be 
obtained if the Sgr dSph is in fact at $D \approx 24\ \kpc$, i.e., 14\% 
closer than adopted by LM10 based on the \citet{sie11} \hst\ 
observations of the Sgr core. This effect will be demonstrated below
when we compare our PM measurements with the $N$-body models of
\citet{pen10}. A lower distance would be consistent 
with some values that have been proposed and used previously in the 
literature \citep[e.g.,][]{law05,bel06}. 

Possibly the LM10 adopted solar velocity is somewhat too low
{\it and} the adopted Sgr dSph distance is somewhat too high. Or
alternatively, other effects may contribute. This could be
explored explicitly through future $N$-body models that are
specifically tailored to fit the new \hst\ PM data.

There is also a (smaller) mismatch between the PM data and the LM10
model for the leading stream stars in {\bf FIELD 1} (green data points
in Figure~\ref{fig:pmresults}). In particular, their observed $\mu_W$
component is somewhat smaller in the data than in the models. This
mismatch is also evident in Figure~\ref{fig:field1}c. It is likely
that an understanding of this mismatch will provide improved insights
into stream models. 

The measured dispersions for the fields are indicated with horizontal
bars above and below the data points in Figure~\ref{fig:pmresults}. 
By-and-large, these dispersions are of similar size as the PM spreads 
for the LM10 model particles (with the possible exception of 
{\bf FIELD 3}). This suggests that much of the spreads in our PM 
measurements are reflective of intrinsic scatter between individual 
stream star PMs, and are not merely due to PM measurement errors. 
In turn, this implies that our PM uncertainties are small enough to be 
able to probe the internal kinematics of the stream, and not merely 
its bulk motion. It is intriguing in this context that in three of 
our four fields, we find indications that two distinct kinematical 
components (A and B) may exist within the same arm and wrap of the 
stream. This is not predicted by the LM10 model, and may tell us 
something new about the structure of the Sgr dSph prior to its 
disruption. Possibly related, it is of interest to note that there 
does also exist a spatial bifurcation in the leading arm of the stream 
\citep{bel06} which is not currently well understood.

\begin{figure*}
\epsscale{1.15}
\plotone{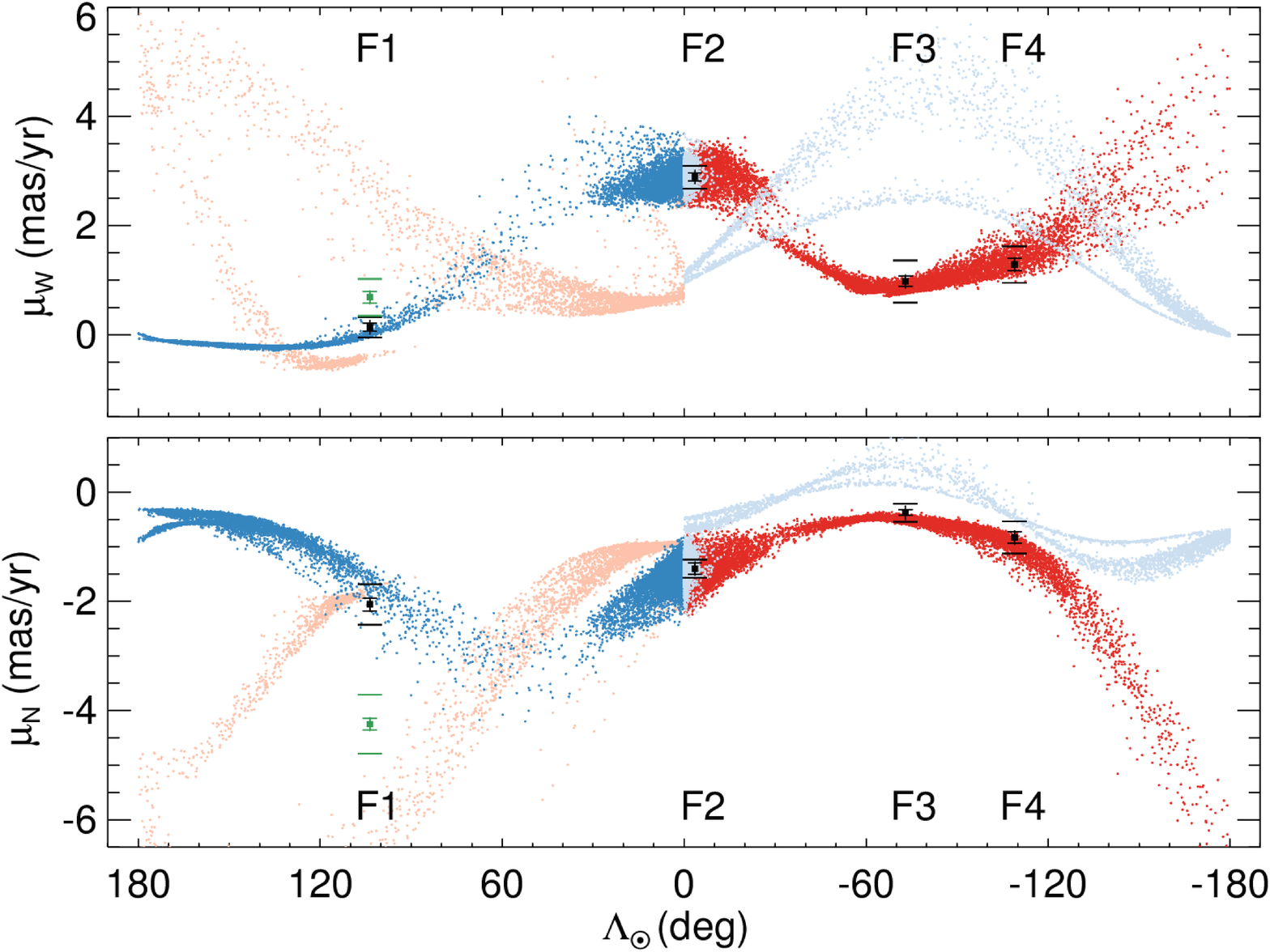}
\caption{Same as Figure~\ref{fig:pmresults} but comparison with 
         $N$-body models of \citet{pen10}.
\label{fig:pmresults_pena}}
\end{figure*}

We have also compared our PM measurements with the $N$-body models of 
\citet{pen10} in a similar manner as we did for LM10. The result is 
shown in Figure~\ref{fig:pmresults_pena}. Again, the variation of the
\hst\ PM measurements with Sgr longitude shows good overall qualitative
agreement with the $N$-body model predictions. This provides yet
another important consistency check on our PM measurements. It also
confirms most of what we already concluded from the comparison with
the LM10 model. However, there are two differences between the LM10
and \citet{pen10} data-model comparisons that are worth noting. On the 
one hand, our average PM for {\bf FIELD 2} is better matched
by the Pe\~{n}arrubia et al. model than by the LM10 model. This is
probably because \citet{pen10} used a distance to the Sgr dSph of only 
$D = 25 \kpc$ (smaller than the $D = 28 \kpc$ adopted by LM10), but 
with a similar Solar motion. On the other hand, the agreement between 
the PM data and the $N$-body predictions for the leading stream stars 
in {\bf FIELD 1} is significantly worse for the \citet{pen10} model 
than for the LM10 model.

In Paper~II we quantitatively compare the new \hst\ data to Sgr 
stream models. We use this comparison to shed new light on topics 
such as the structure and distance of the Sgr stream, the solar 
velocity in the MW disk, and the shape of the MW's dark halo.

%
%

\section{Summary and Conclusions}
\label{sec:conclusions}

The Sgr stream is an ideal target for probing in detail several
important topics, including the tidal disruption of dwarf galaxies,
the hierarchical buildup of stellar halos, the shape, orientation and
mass of the MW's dark halo, and the velocity of the Sun in the MW
disk. To be able to better use the stream to address these issues, we
have presented an \hst\ study of the variation in stellar PMs along
the stream. 

We observed four fields along the Sgr stream: one field in the
trailing arm, one field near the Sgr dSph tidal radius, and two fields
in the leading arm. These fields span a range of more than 200 degrees
along the Sgr stream. We combined the results with existing archival
data to yield time baselines of 6--9 years. From the data we
determined two-band photometry and absolute PMs for individual stars
in these fields. We used distant background-galaxies to define a
stationary reference frame. Our sophisticated PM measurement
techniques have been previously developed, tested, and applied in the
context of other Local Group projects. Any residual systematic PM
errors are expected to be below the level of the random measurement
errors.

By combining CMD and PM information it is possible to identify
individual Sgr stream stars in each of the fields, and to reliably
reject most foreground and background objects. This yields for
the first time accurate individual (as opposed to average) PM
measurements of stars in the stream. This allows separation of
different kinematical components and wraps within the stream. For
example, in our trailing-arm field, the PMs allow us to kinematically
separate ten trailing-arm stars from three leading-arm stars. These
leading-arm stars are a full revolution ahead in their orbit around
the MW compared to the trailing-arm stars.

We compared our results to the small body of existing average PM
measurements for the Sgr stream and dSph. Our results are broadly
consistent with the existing work, but have higher accuracy. Moreover,
our results provide the first PM measurements in the leading arm of the
stream, which can help constrain MW halo properties over a wider range
of radii.

For a preliminary assessment of what the new \hst\ PM data may imply 
for our understanding of the Sgr stream, we qualitatively compared the
data to the PM predictions of the $N$-body models of LM10 and 
\citet{pen10}. Overall, our measurements follow quite closely the 
predicted PM trends with Sgr longitude. Also, the measured PM 
dispersions are broadly comparable to those in the models. This 
provides a successful consistency check on the PM measurements. It
also constitutes a remarkable success for the $N$-body models, which
were fit only to distance and LOS velocity data for the stream, with
no reference to PMs.

Nonetheless, there are some areas of mismatch between the models and
our PM data, as discussed in Section~\ref{sec:PMmodel}. Also, in three 
of our fields we find indications that two distinct kinematical 
components (A and B) may exist within the same arm and wrap of the 
stream. This may tell us something new about the structure of the Sgr 
dSph prior to its disruption. In Paper II we will quantitatively 
compare the new \hst\ PM data to Sgr stream models, and we use this 
comparison to shed new light on topics such as the structure and 
distance or the Sgr stream, the solar velocity in the MW disk, and the 
shape of the MW?s dark halo.

The observational techniques presented here will be equally useful for
the study of other structures in the MW halo. For example, we have an
ongoing \hst\ observing program (GO-13443, PI: R.~P.~van der Marel) to
determine PMs along the Orphan Stream. This will better constrain its
orbit, and may allow identification of its progenitor (if it is not 
already entirely disrupted). When observations from different streams 
are combined, it should also be possible to constrain more tightly the
shape and mass of the MW dark halo. We also have an ongoing
\hst\ archival legacy program (AR-13272, PI: R.~P.~van der Marel), to
determine the PMs of metal-poor halo stars in random pointings
multiply-imaged by \hst. This will not only constrain the velocity
dispersion anisotropy of the dynamically hot halo, buy may also lead
to identification of new streams if cold structures are found to exist
in PM space.

\acknowledgments

We thank the referee for constructive feedback that helped improve the 
presentation of our results. Annie Robin kindly provided updated Besancon 
models for our target fields. Jorge Pen\~{n}arrubia kindly provided his 
$N$-body models.
Support for this work was provided by NASA through a grant for program
GO-12564 from the Space Telescope Science Institute (STScI), which is
operated by the Association of Universities for Research in Astronomy
(AURA), Inc., under NASA contract NAS5-26555. This research has made
use of the NASA/IPAC Extragalactic Database (NED) which is operated by
the Jet Propulsion Laboratory, California Institute of Technology,
under contract with the National Aeronautics and Space Administration.
J.L.C. acknowledges support from NSF grants AST 09-37523 and 
AST 14-09421.

{\it Facilities:} \facility{\hst\ (ACS/WFC)}.

\appendix 

\section{A. PSF Kernels}

To determine accurate PMs, it is important to account for PSF
differences between epochs. Specifically, it is known that a subtle
change in the ACS/WFC PSF was introduced by the Space Shuttle
Servicing Mission 4. We account for this through application of a
7$\times$7 pixel convolution kernel to one of the two epochs. As
discussed in Section~\ref{sec:starastrom}, our {\bf FIELDS 1}, {\bf
  3}, and {\bf 4} are too sparse to reliably derive these kernels from
the actual point sources observed in these fields. So for these fields
we used an alternative method based on library PSFs.

We first created two sets of simulated images, representing the first
and second epoch, by injecting 1,000 stars with random brightness
($S/N > 100$) in random locations. The background characteristics of
the simulated images, such as the mean and standard deviation, were
chosen to follow those of our observed images. Position-dependent
library PSFs \citep{and06} constructed separately for the pre- and
post-SM4 period were used when injecting stars into the simulated
first- and second-epoch images, respectively. Using these stars in the
two sets of simulated images, we obtained a kernel that accounts for
the differences between the pre- and post-SM4 PSFs, similarly to our
analysis of the {\bf FIELD 2} data (see Section~\ref{sec:starastrom}).
We did this separately for each of the F775W and F814W filters. These
kernels were then applied to the templates when fitting a star or
galaxy in the second-epoch images for {\bf FIELDS 1}, {\bf 3}, and
{\bf 4}.

\section{B. Absolute Proper Motion of the Globular Cluster NGC~6652}

In the course of our PM analysis of {\bf FIELD 2}, we have also
obtained the absolute PM of the globular cluster NGC~6652:
\begin{equation}
\label{PM_NGC6652}
  \mu_{\rm NGC6652} = (\muw, \mun) =
  (5.66 \pm 0.07, -4.45 \pm 0.10)\ {\rm mas\ yr}^{-1} .
\end{equation}
To determine the implied Galactocentric motion, we adopt a Cartesian
Galactocentric coordinate system ($X, Y, Z$), with the origin at the
Galactic Center, the $X$-axis pointing in the direction from the Sun
to the Galactic Center, the $Y$-axis pointing in the direction of the
Sun's Galactic rotation, and the $Z$-axis pointing towards the
Galactic North Pole. The position and velocity of an object in this
frame can be determined from the observed sky position, distance,
line-of-sight velocity, and proper motion.

For the distance $R_0$ of the Sun from the Galactic Center and 
the circular velocity of the local standard of rest (LSR), we adopt 
the recent values of \citet{mcm11}: $R_{0} = 8.29 \pm 0.16\ \kpc$ and 
$V_{0} = 239 \pm 5\ \kms$. For the solar peculiar velocity with 
respect to the LSR we adopt the estimates of \citet{sch10}: $(U_{\rm pec}, 
V_{\rm pec}, W_{\rm pec}) = (11.10, 12.24, 7.25)\ \kms$ with 
uncertainties of $(1.23, 2.05, 0.62)\ \kms$.

The distance to NGC~6652 is $10.5 \pm 0.5$ kpc \citep{cha00}, and 
the heliocentric line-of-sight velocity is $-111.7 \pm 5.8\ \kms$
\citep[][2010 edition]{har96}. These imply a Galactocentric 
$(X, Y, Z)$ position 
\begin{equation}
  {\vec r} = (2.0, 0.3, -2.1)\ \kpc ,
\end{equation}
and a Galactocentric velocity vector 
\begin{equation}
  {\vec v} = (-58.0 \pm 6.2, -70.6 \pm 16.4, 187.4 \pm 8.2)\ \kms .
\end{equation}
The corresponding Galactocentric radial and tangential velocities are
\begin{equation}
  (V_{\rm rad}, V_{\rm tan}) = 
   (-181.0 \pm 7.2, 103.4 \pm 31.9)\ \kms ,
\label{vradtaneq}
\end{equation}
and the observed total velocity of NGC~6652 with respect to the MW is
\begin{equation}
  v \equiv |{\vec v}| = 208.5 \pm 11.8\ \kms .
\label{vtoteq}
\end{equation}
The listed uncertainties above are obtained from a Monte-Carlo scheme
that propagates all observational distance and velocity uncertainties
and their correlations, including those for the Sun.
\\
\\

\end{document}